\documentclass[authoryear]{elsarticle}

\usepackage{soul}

 \usepackage[utf8]{inputenc} 
\usepackage[english]{babel}
\usepackage{fancyhdr} 
\usepackage{url} 
\usepackage{subcaption} 
\usepackage{natbib} 
\usepackage{multirow} 
\usepackage{float}
\floatstyle{plaintop}
\restylefloat{table}

\usepackage{amssymb}
\usepackage[intlimits]{amsmath} 
\usepackage{amsthm} 
\usepackage[german]{nomencl}
\usepackage{booktabs} 
\usepackage{geometry}
\usepackage{enumitem}
\usepackage{multicol}
\usepackage[table]{xcolor}
\usepackage{bbm}

\usepackage{tikz}
\usepackage{tikz-qtree,tikz-qtree-compat}
\usetikzlibrary{positioning}

\makenomenclature
\usepackage[pdfborder={000},colorlinks=true,linkcolor=blue,citecolor=blue]{hyperref}
\usepackage{listings}
\usepackage[all]{xy}
\usepackage{color}

\makeatletter
\def\namedlabel#1#2{\begingroup
    #2%
    \def\@currentlabel{#2}%
    \phantomsection\label{#1}\endgroup
}
\makeatother

\definecolor{b}{rgb}{0,0,.8}	
\definecolor{g}{rgb}{0,.6,0}	
\definecolor{n}{rgb}{0,0,0}	
\definecolor{h}{rgb}{0.4,0.2,0.2}	
\definecolor{v}{rgb}{0.2,0.6,0}

\renewcommand{\AA}{{\mathcal{A}}}
\newcommand{\BB}{{\mathcal{B}}}

\newcommand{\DD}{{\mathcal{D}}}

\newcommand{\HH}{{\mathcal{H}}}

\newcommand{\QQ}{{\mathcal{Q}}}

\newcommand{\TT}{{\mathcal{T}}}

\newcommand{\WW}{{\mathcal{W}}}

\newcommand{\bsX}{\boldsymbol X}

\newcommand{\bsbeta}{\boldsymbol \beta}

\newcommand{\eps}{{\varepsilon}}

\DeclareMathOperator*{\argmin}{arg\,min}

\newcommand{\ov}\overline
\newcommand{\what}{\widehat}

\newcommand{\rig}\right
\newcommand{\lef}\left
\newcommand{\nf}\normalfont

\definecolor{gray}{rgb}{0.5,0.5,0.5}
\definecolor{red}{rgb}{0.8,0,0}
\definecolor{dred}{rgb}{0.5,0,0}
\definecolor{cyan}{rgb}{0,0.7,.2}
\definecolor{dcyan}{rgb}{0,0.5,.5}

\definecolor{strikeout}{rgb}{0.5,0.5,.5}
\definecolor{newchange}{rgb}{0.8,0,0}

\begin{document}

\title{Quantile Regression for Qualifying Match of \\ GEFCom2017 Probabilistic Load Forecasting
}

 \author{Florian Ziel\corref{cor1}}
 \ead{florian.ziel@uni-due.de}
 \address{House of Energy Markets and Finance, University of Duisburg-Essen, Germany}

\journal{International Journal of Forecasting}

\begin{keyword}
Load forecasting \sep Probabilistic forecasting \sep Quantile regression \sep Periodic pattern  \sep Seasonal interaction \sep Long-term trend \sep GEFCom
\end{keyword}
\begin{frontmatter}
\lhead{\nouppercase{\leftmark}}
\begin{abstract}

We present a simple quantile regression-based forecasting method that was applied in a probabilistic load forecasting framework of the Global Energy Forecasting Competition 2017 (GEFCom2017). The hourly load data is log transformed and split into a long-term trend component and a remainder term. The key forecasting element is the quantile regression approach for the remainder term that takes into account weekly and annual seasonalities such as their interactions. Temperature information is only used to stabilize the forecast of the long-term trend component. Public holidays information is ignored. Still, the forecasting method placed second in the open data track and fourth in the definite data track with our forecasting method, which is remarkable given simplicity of the model. The method also outperforms the Vanilla benchmark consistently.



\end{abstract}
\end{frontmatter}


\section{Introduction}

The Global Energy Forecasting Competition 2017 (GEFCom2017) is an international
forecasting challenge on probabilistic load forecasting. It is the successor of the second GEFCom2014, see 
\cite{hong2016probabilistic}.
The GEFCom2017 was a two stage competition with 
a qualifying match  and a 
final match which contained different tasks. 

This paper presents the forecasting methodology developed by one of the winning teams, \emph{'simple\_but\_good'}, during the qualifying match which involved about 100 participating teams. 
We explain briefly the competition settings and rules and introduce the underlying data. 
Then, we present the forecasting model which is essentially a quantile regression model. Afterwards, we show and discuss the obtained  results. 

Quantile regression elements are popular in probabilistic load and price forecasting literature, as in e.g.  \cite{Haben20161017}, \cite{gaillard2016additive}, 
\cite{maciejowska2016hybrid}, \cite{taieb2016forecasting} and \cite{liu2017probabilistic}.
These elements are especially useful in situations where forecasting evaluation is based on quantile rules.

\section{Relevant Competition Setting}

\begin{figure}[htb!]
\centering
  \resizebox{.99\textwidth}{!}{
\begin{tikzpicture}[level 1/.style={level distance=1.5cm}]
\Tree
[.TOTAL 
  [.ME  ] [.NH ] [.VT  ] [.CT ] [.RI  ] [.MA 
                      [.MA.SE ] 
                      [.MA.WC ] 
                      [.MA.NE  ]
                   ] 
]
\end{tikzpicture} 

\begin{tabular}{r|l|l}
Task& Subm. due date &  Forecasting window \\ \hline
1 & 15 Dec 2016& 1-31 Jan 2017 \\
2 & 31 Dec 2016& 1-28 Feb 2017 \\
3 & 15 Jan 2017& 1-28 Feb 2017 \\
4 & 31 Jan 2017& 1-31 Mar 2017 \\
5 & 14 Feb 2017& 1-31 Mar 2017 \\
6 & 28 Feb 2017& 1-30 Apr 2017
 \vspace{29mm}
\end{tabular}

}
\vspace{-29mm}

\caption{The hierarchical structure (left) of the zones ISO New England (TOTAL), Maine (ME), New Hampshire (NH), Vermont (VT), Connecticut (CT), Rhodes Island (RI) and Massachusetts (MA)
 with its three subregions MA.SE, MA.WC and MA.NE. The competition schedule (right) of all 6 tasks.  
 }
\label{fig_hira}
 \end{figure}
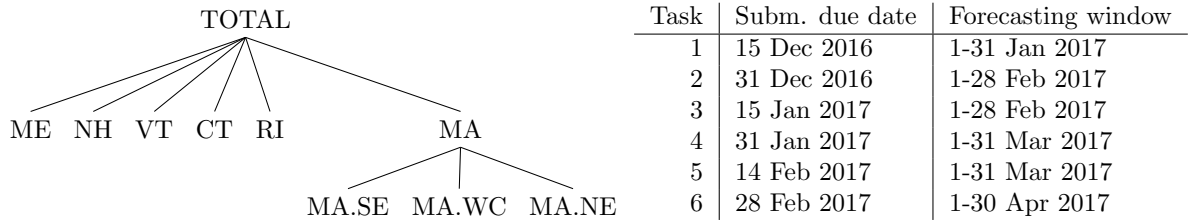

The GEFCom2017 is 
probabilistic hierarchical load forecasting competition. 
The main task of the qualifying match was to forecast the electricity load 
in the ISO New England zone in real time. 
 In particular, the participants had to make predictions as to the electricity demands of 10 zones located within ISO New England. These zones follow a specific hierarchical structure, which is visualized in Figure \ref{fig_hira}.
There are 8 zones on the lowest levels: Maine (ME), New Hampshire (NH), Vermont (VT), Connecticut (CT), Rhodes Island (RI) and three zones in Massachusetts (MA.SE, MA.WC and MA.NE). 
Then, are two additional zones, Massachusetts (MA) itself which contains the three sub-zones and the ISO New England zone (TOTAL). 

 The qualifying match had a defined-data track and an open-data track. 
In the latter one all information the forecaster managed to find could be used to create forecasts. 
The former limited the scope of data allowed for making a prediction to past load and temperature (dry-bulb and dew-point) data as published by ISO New England such as 
public holiday information.\footnote{New Year’s Day, Birthday of Martin Luther King, Jr., Washington’s Birthday, Memorial Day, Independence Day, Labor Day, Columbus Day, Veterans Day, Thanksgiving Day, Christmas Day}


Each track contains in total six tasks with medium term forecasting horizons. The respective detailed schedule 
is given in Figure \ref{fig_hira}, 
 the more thorough description of the content of the tasks is given in what follows.
Since ISO New England publishes usually the data for the previous month prior to the 15th day of each month, the most recently available data for task 1 with January's load was November's load.
Thus, the settings of the tasks 2 and 3 as well as  4 and 5 are not identical even though forecasting window is. 
 Finally, the final task 6 got double weights in the final evaluation. 
Having completed the exercises, the respective code and a report had to be submitted confidentially to the organizer, Tao Hong.

 The competition focuses on probabilistic forecasting and pays a particular attention to 
 quantile forecasting. 
More specifically, we require the quantile forecasts of $\QQ  = \{0.1,0.2, \ldots, 0.9\}$ to minimize the competition evaluation score function.
The score function which is used for evaluation is based on the pinball score which can also be referred to as quantile loss. 
For a quantile level $\tau\in \QQ$ the pinball score is given by
\begin{align}
S_{\tau}(y,q)  =  \mathbbm{1}_{ \{ y \geq q \} } \tau (y - q)  +  \mathbbm{1}_{ \{ q > y \} }   (1 - \tau) (q - y) 
=  (y - q) (\tau - \mathbbm{1}_{ \{ y - q < 0 \} } ) 
\label{eq_pb}
\end{align}
 where $q$ is the quantile forecast, $y$ is the observed load value and $\mathbbm{1}$ is the indicator function.
 An intermediate score for each task and for each zone is computed as average pinball score \eqref{eq_pb} over the forecasting horizon and all quantile levels in $\QQ$.

Afterwards, these intermediate scores are normalized (to make zones with different load levels comparable) so that they measure the relative improvements of the forecasting model 
with respect to a benchmark model.
This benchmark model is the Vanilla benchmark that was also used for similar purpose in the previous GEFCom2014, see \cite{hong2016probabilistic}.
Note that the final normalized score is minimized if the pinball scores in \eqref{eq_pb} is minimized for all forecasting horizons, quantile levels, zones and tasks.

\section{Data and descriptive data analysis}

To build a model for the GEFCom2017 we considered hourly load data of all 10 regions. Denote by $\ell_{i,t}$ the electricity load for region $i$ at time $t$. 
Additionally to the load data we also exploit temperature data (DryBulb) $T_{i,t}$ as provided in the competition. 

Prior to the modeling we clock-change adjust the temperature and load data. Hence, $\ell_{i,t}$ and $T_{i,t}$ denotes the already clock-change adjusted data. 
We do standard day-light saving time adjustment for all locations, i.e. we average values of the doubling hour in November and linearly extrapolate values of the missing hour in March.
This gives us 24 observations each day.
For solving all tasks we always consider approximately 10.25 years ($365\times 10 + 92$ days) of in-sample data  for the estimation. 
We also use this set for conducting summary statistics within this paper.

\begin{figure} 
\centering
\includegraphics[width=.24\textwidth]{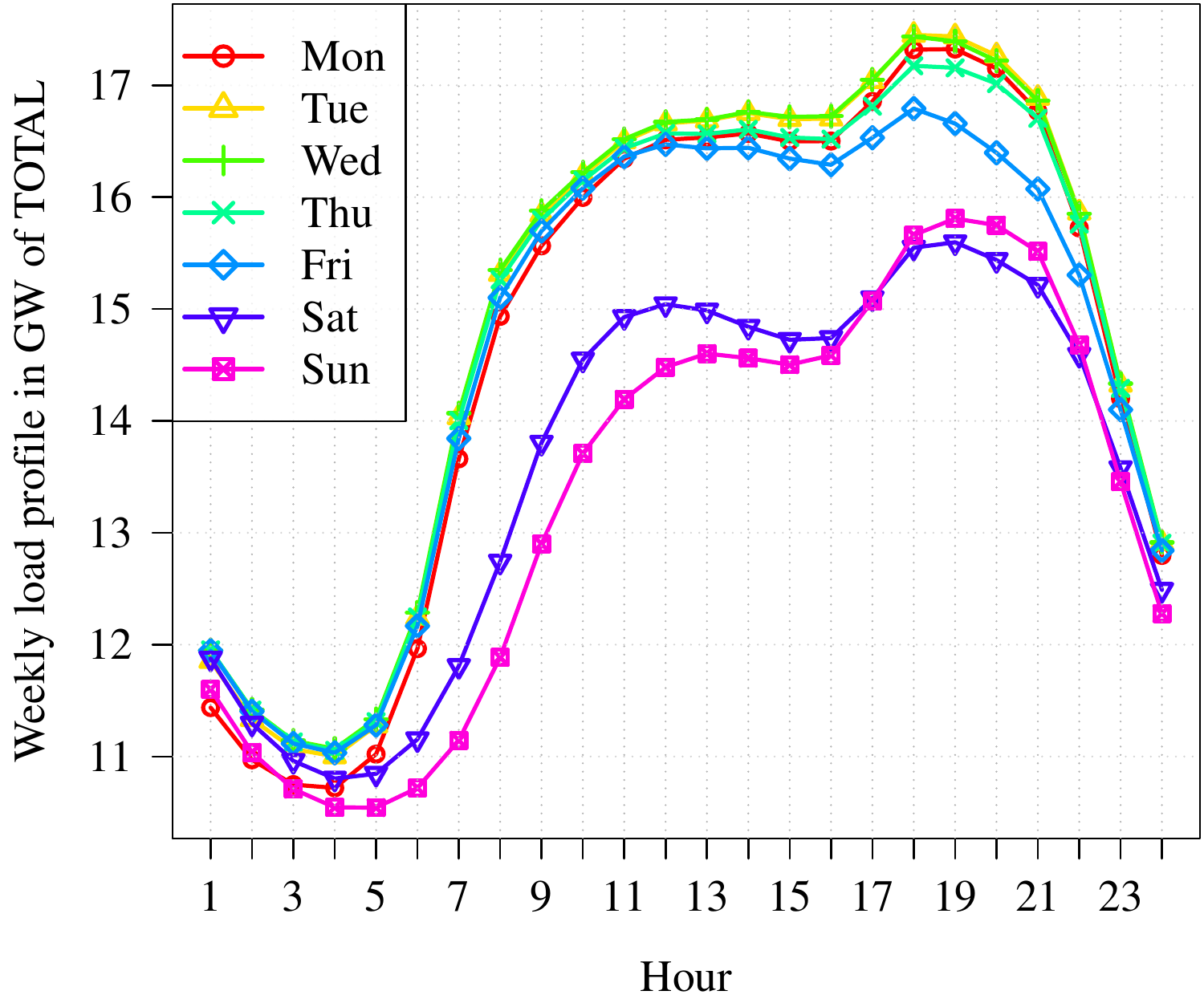} 
\includegraphics[width=.24\textwidth]{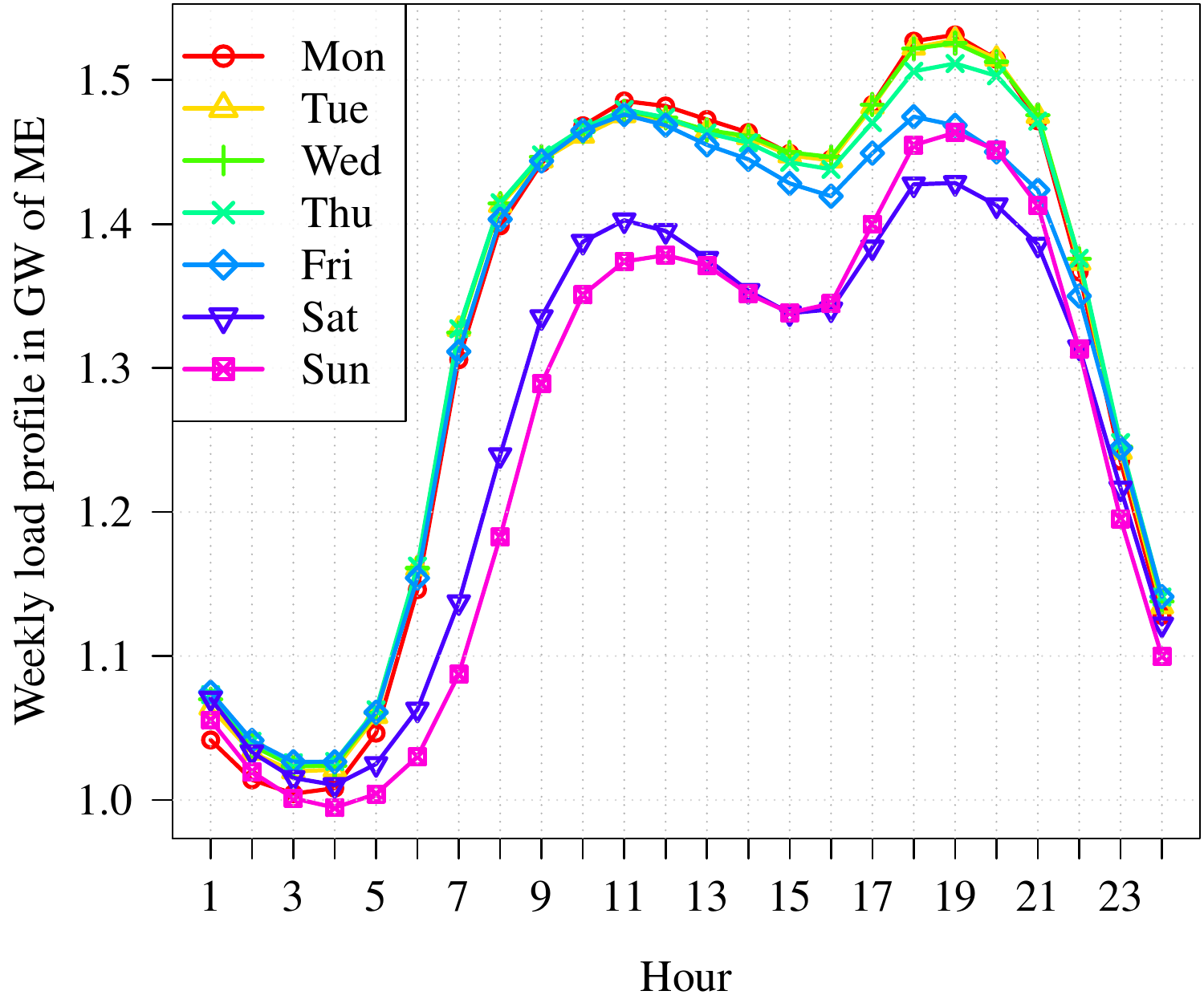} 
\includegraphics[width=.24\textwidth]{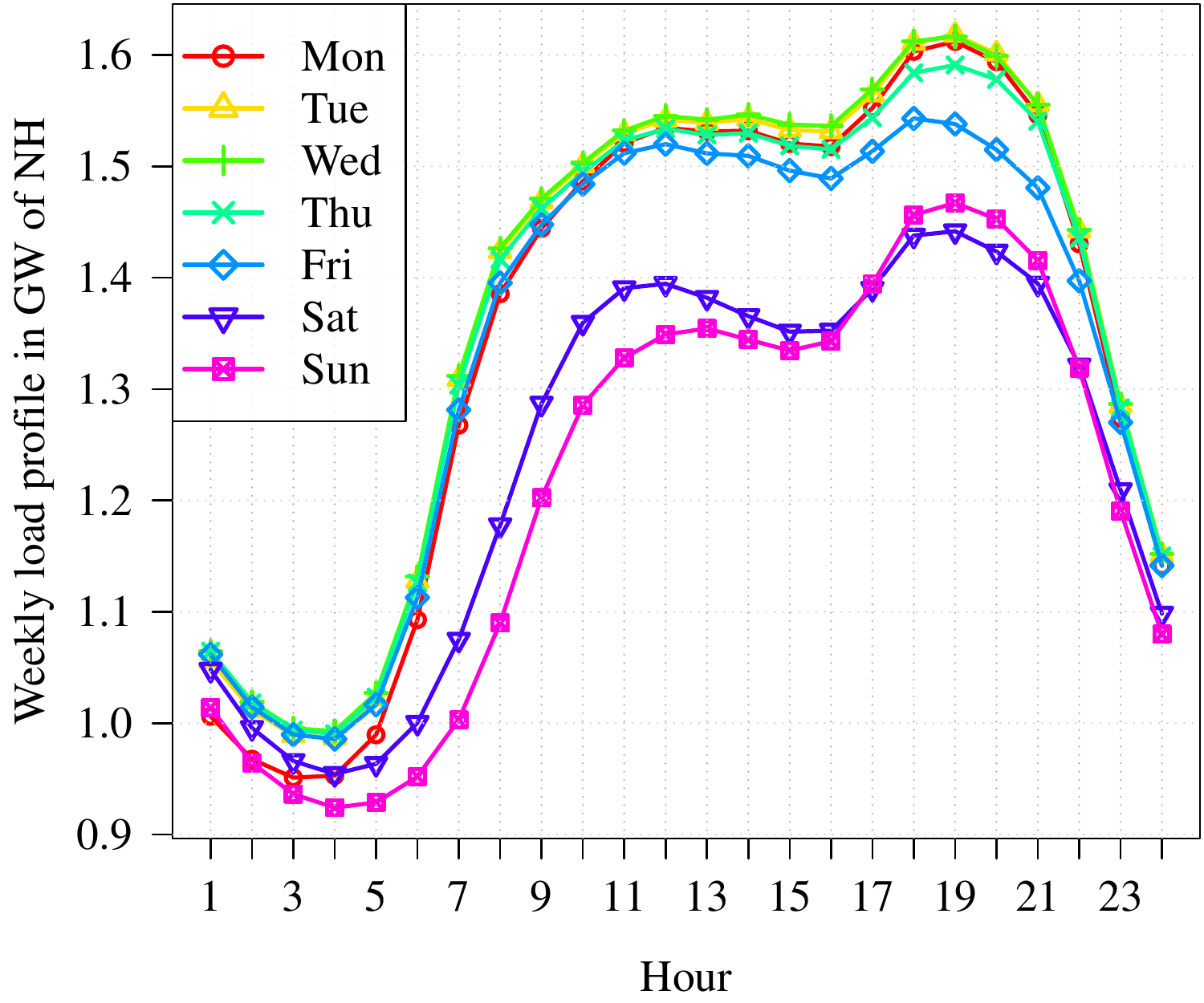} 
\includegraphics[width=.24\textwidth]{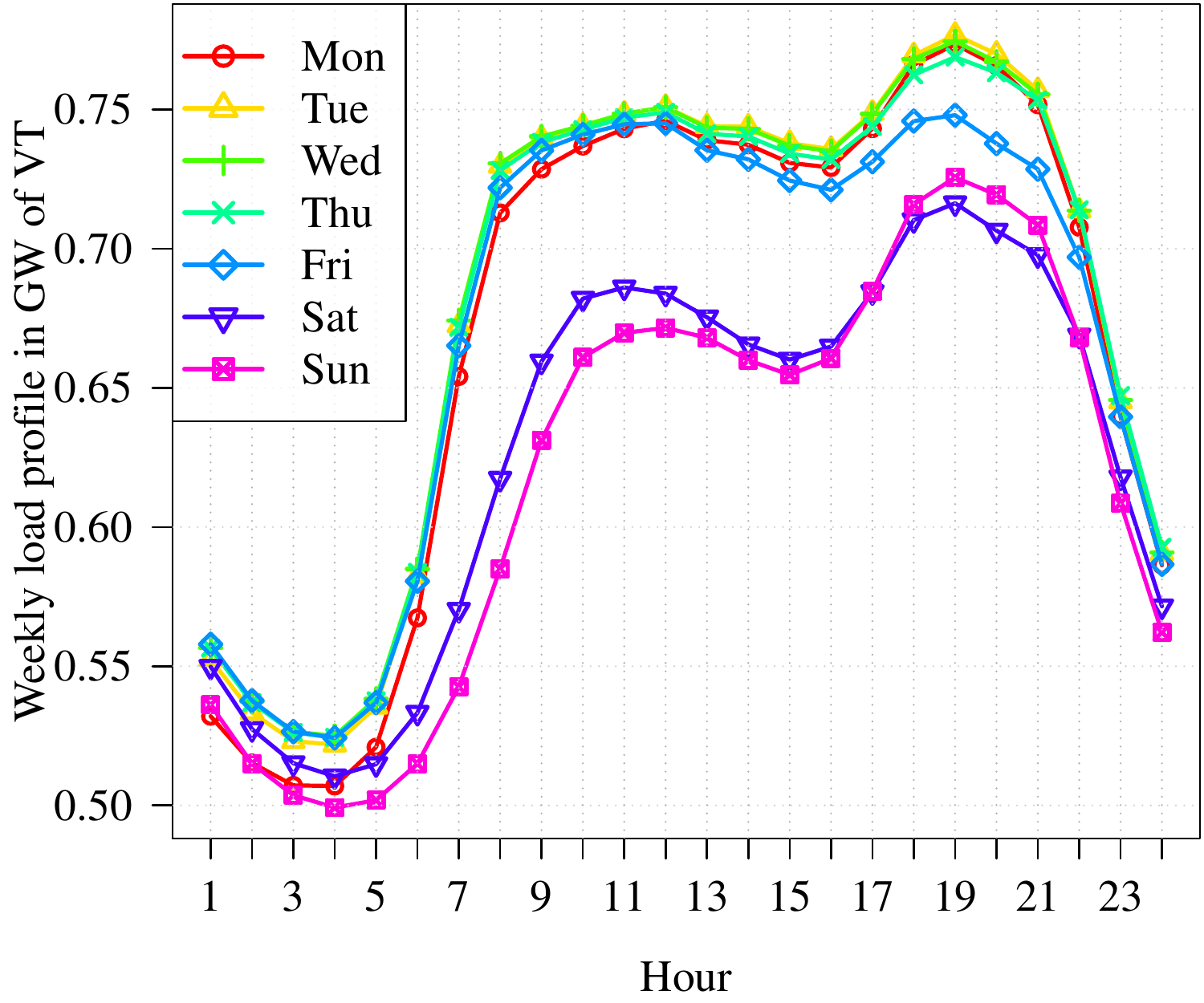} 
\includegraphics[width=.24\textwidth]{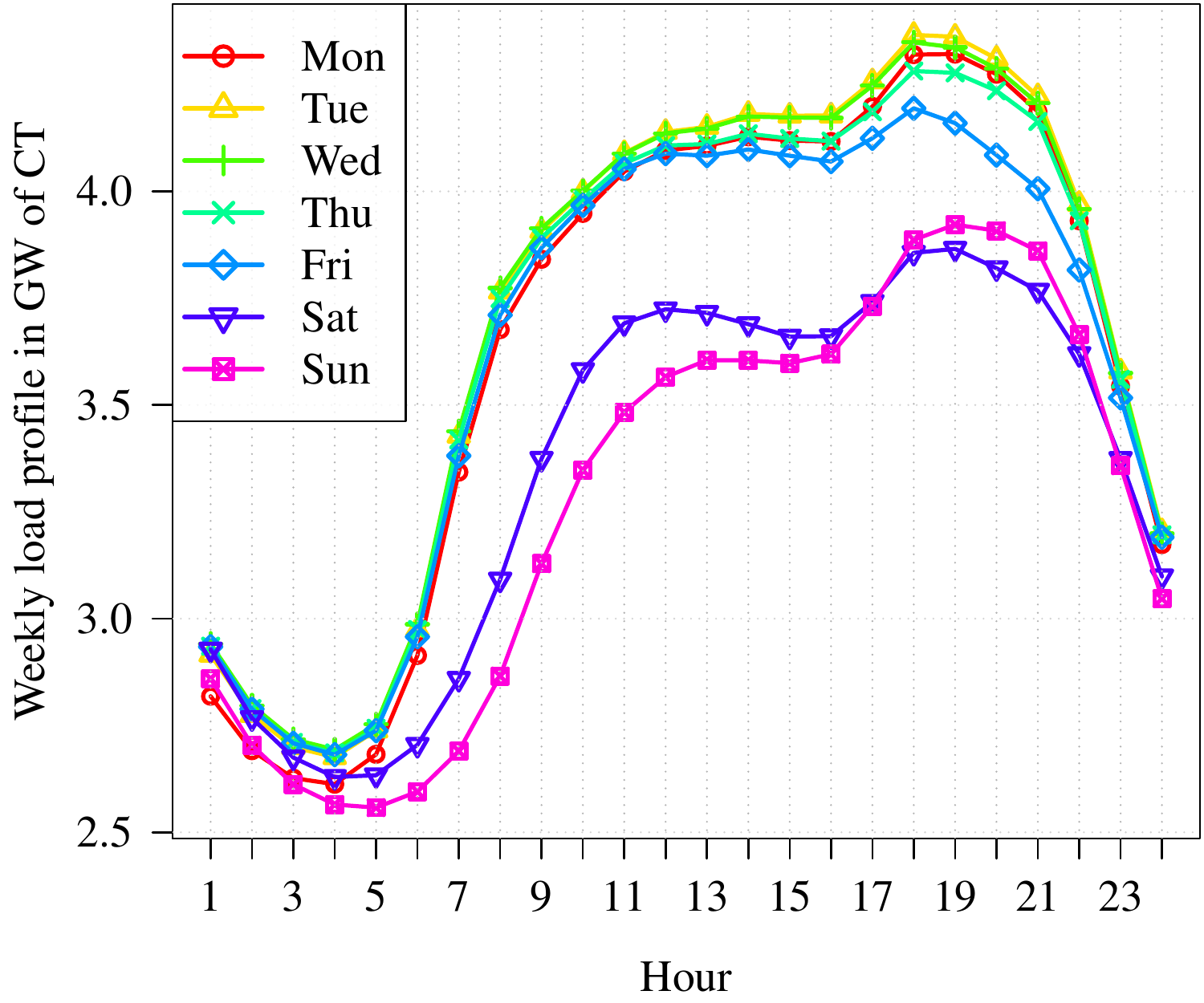} 
\includegraphics[width=.24\textwidth]{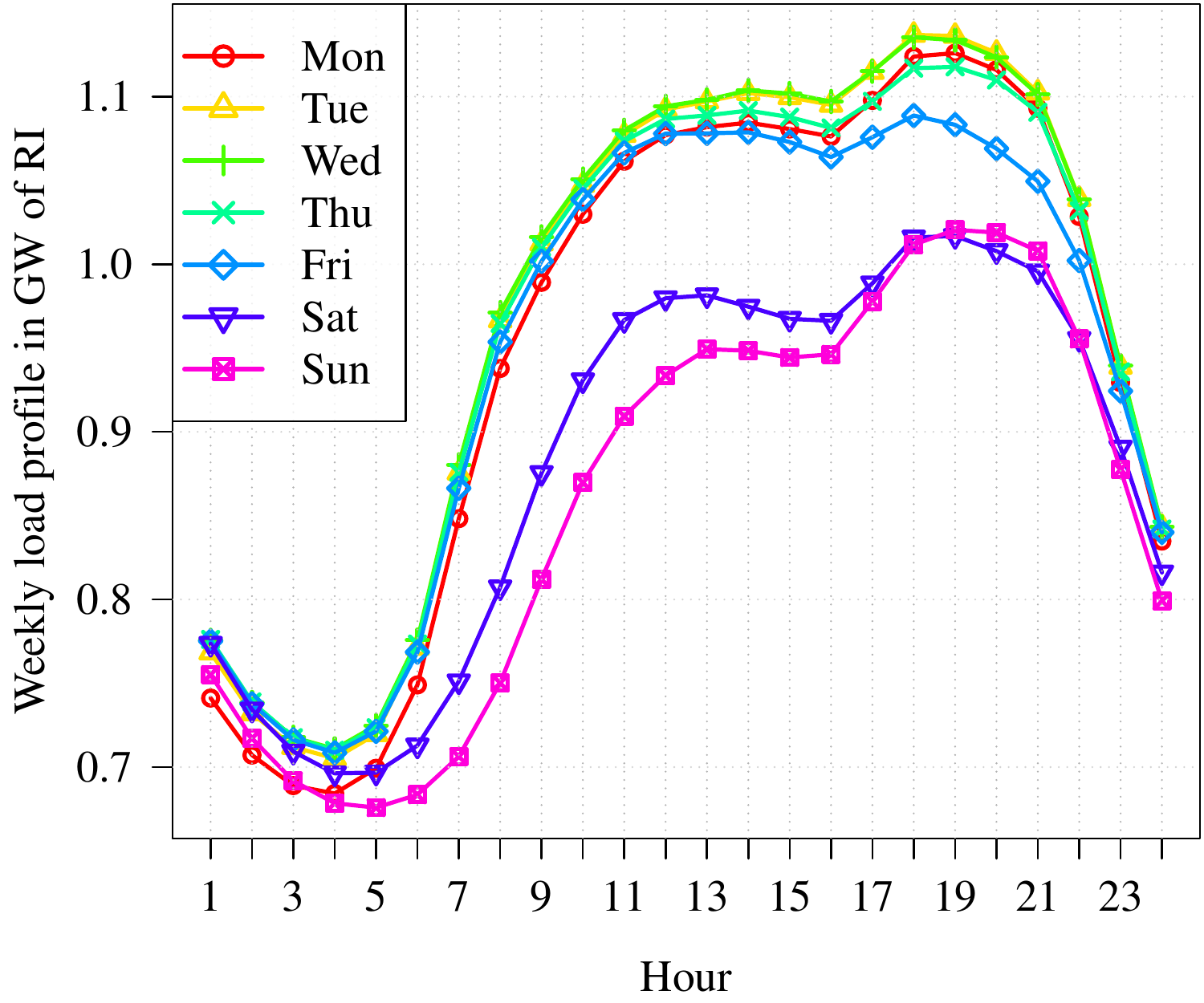} 
\includegraphics[width=.24\textwidth]{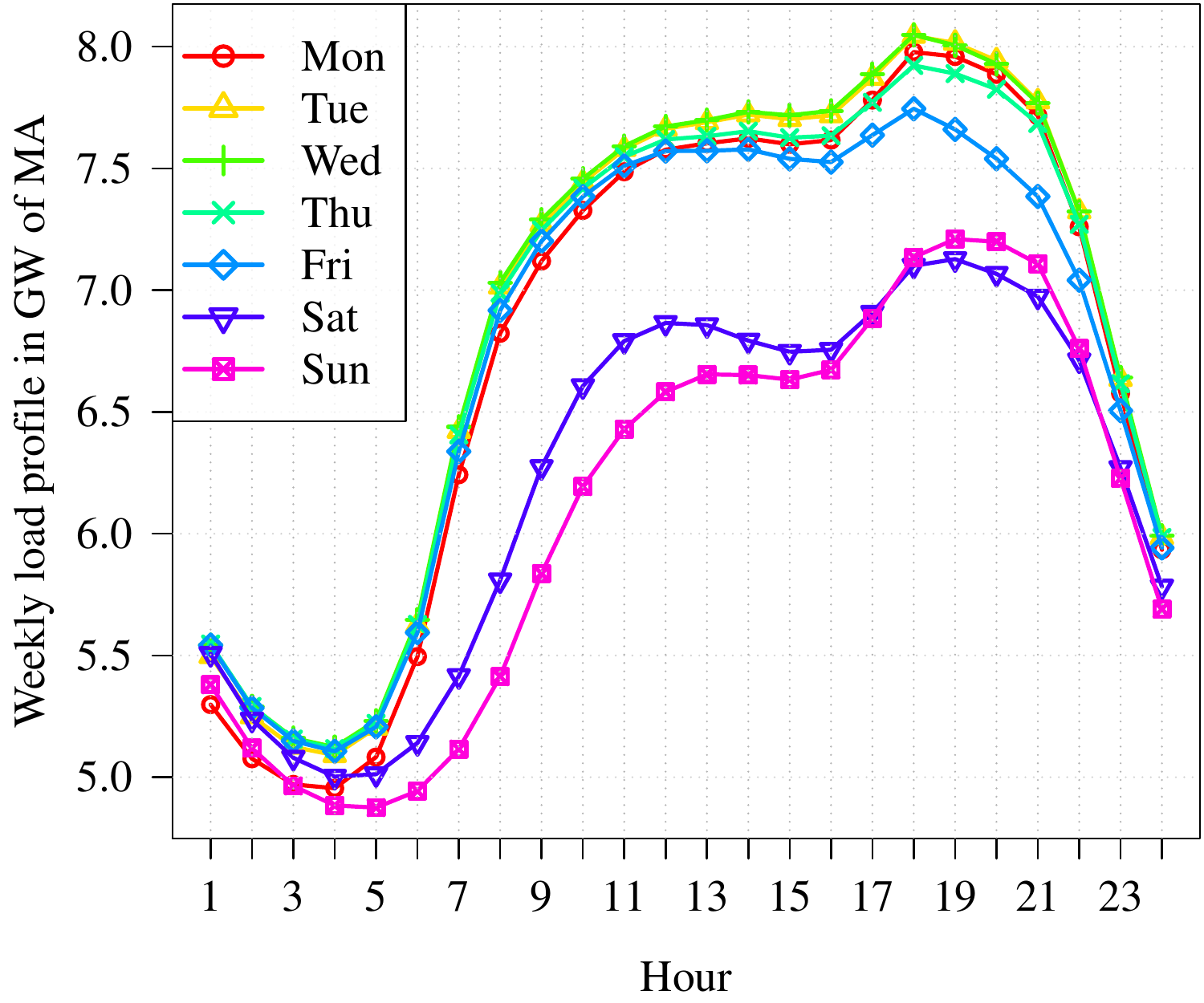} 
\includegraphics[width=.24\textwidth]{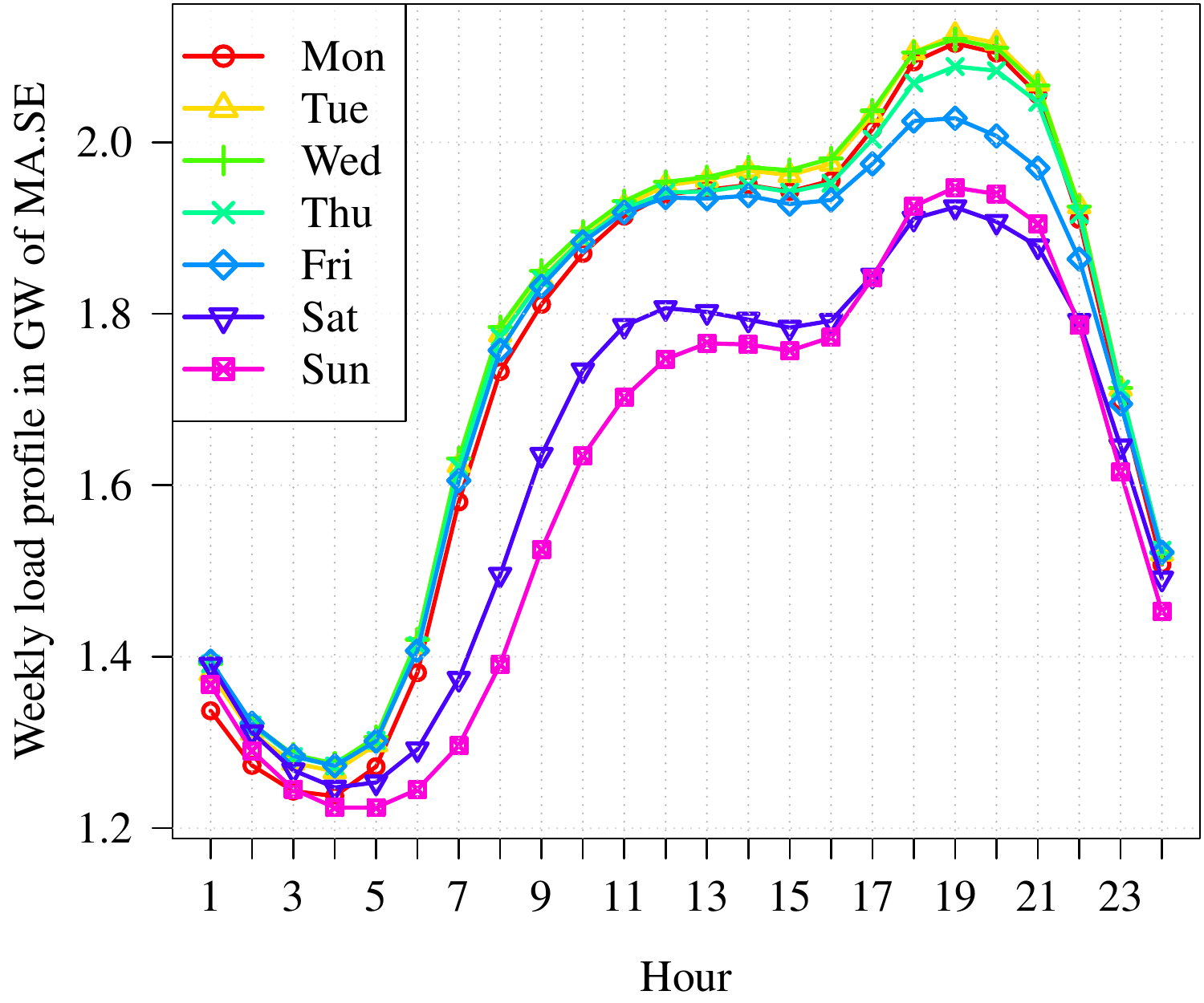} 
\includegraphics[width=.24\textwidth]{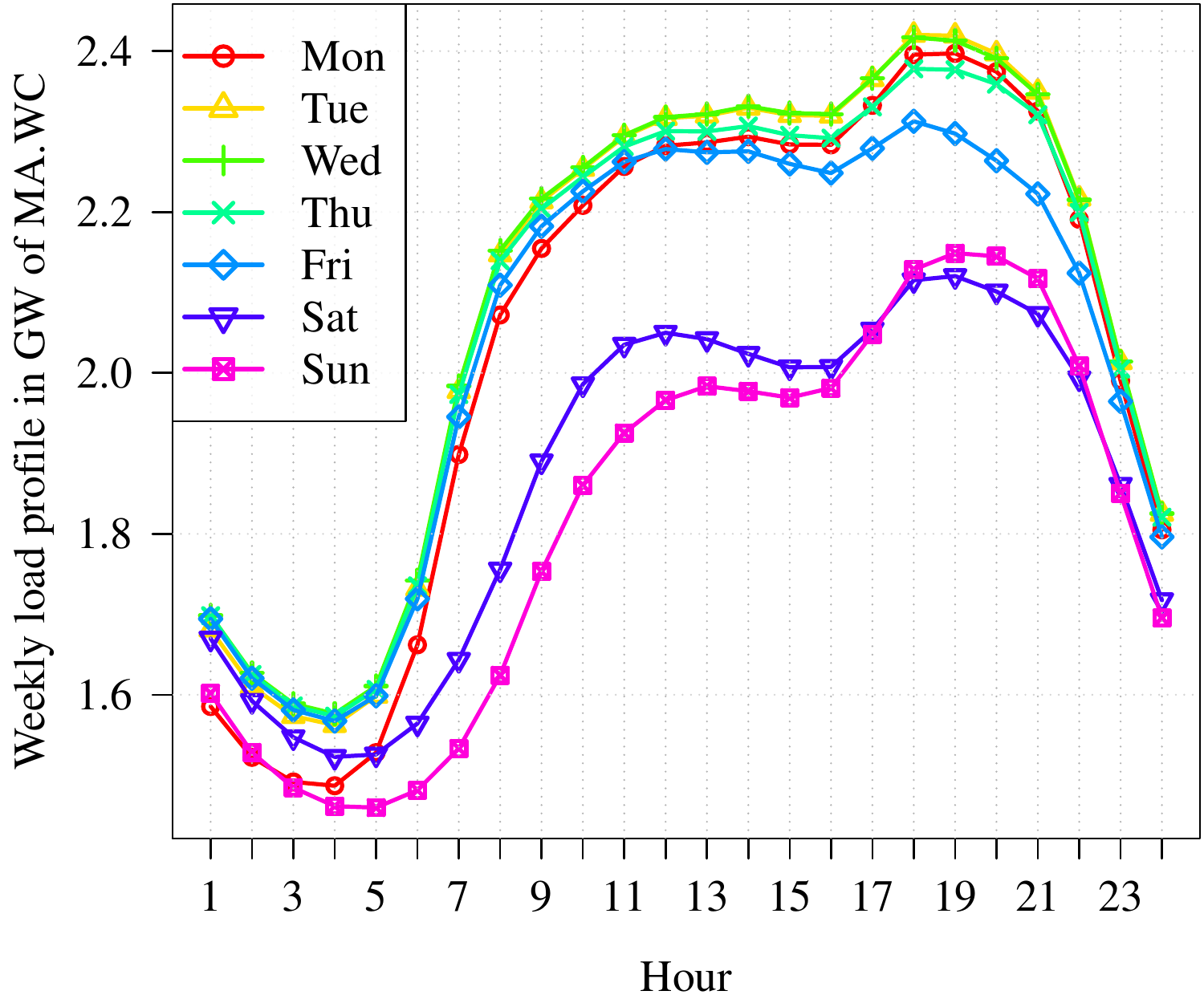} 
\includegraphics[width=.24\textwidth]{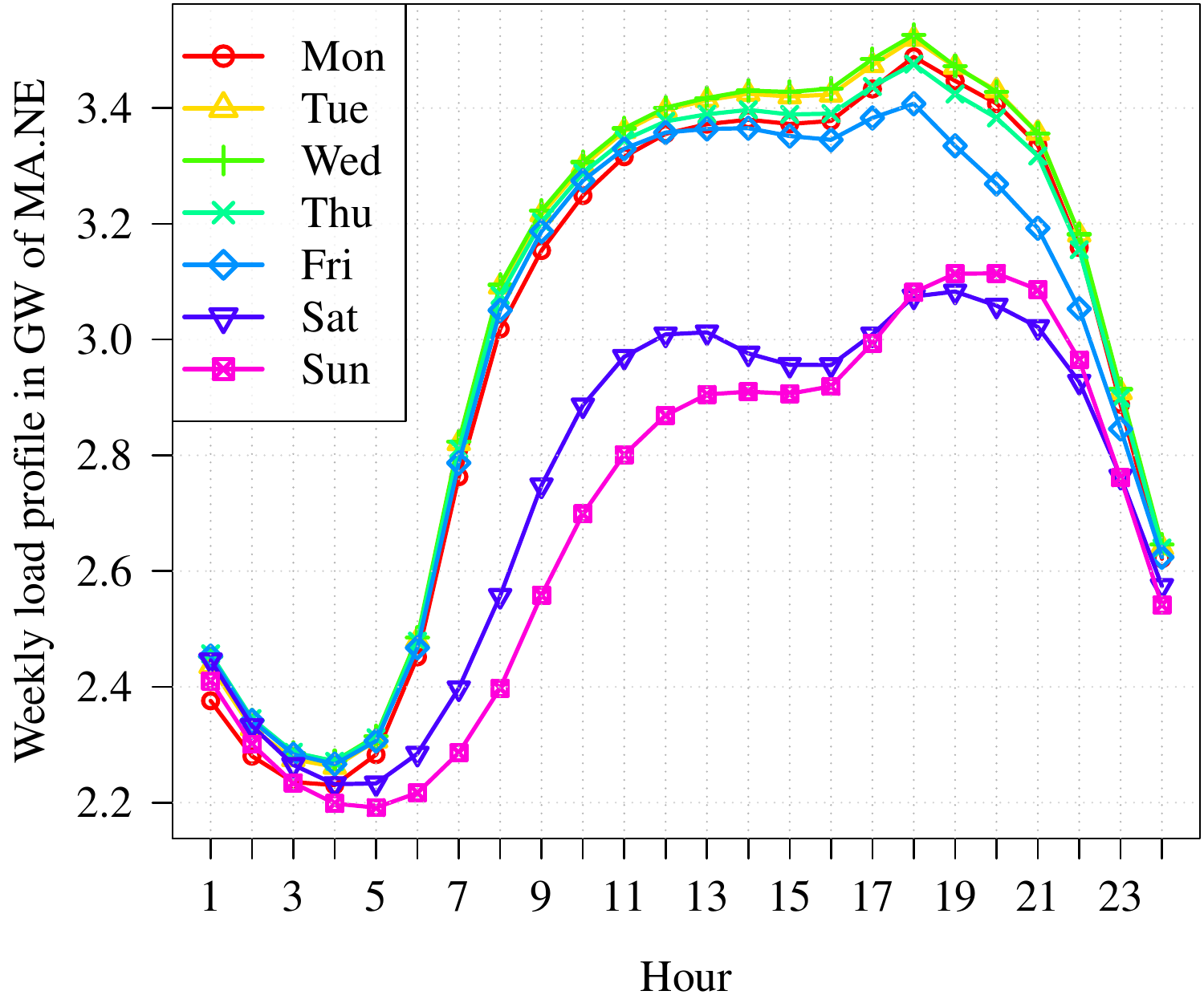} 
\caption{Weekly load profile for all ten zones in consideration in the ordering TOTAL, ME, NH, VT, CT, RI, MA, MA.SE, MA.WC and MA.NE.
}
\label{fig_trend_load_profiles}
\end{figure}

The weekly load profiles of $\ell_{i,t}$ for all ten zones are visualized in Figure \ref{fig_trend_load_profiles}.
We created this Figure by taking sample means of every hour of the week given in the available data set. 
The figure shows that all time series share relatively similar weekly standard behavior. 
Their demand levels are lower during nights and higher during working hours from Monday to Friday.
Furthermore, we also see weekend transition effects on Monday and Friday. As the overall behavior appears to be rather homogeneous in all zones, we consider the same model approach for all of them.
Moreover, we only report and illustrate results for the ISO New England zone (TOTAL) in this paper. 

\begin{figure}[htb!]
\includegraphics[width=.49\textwidth]{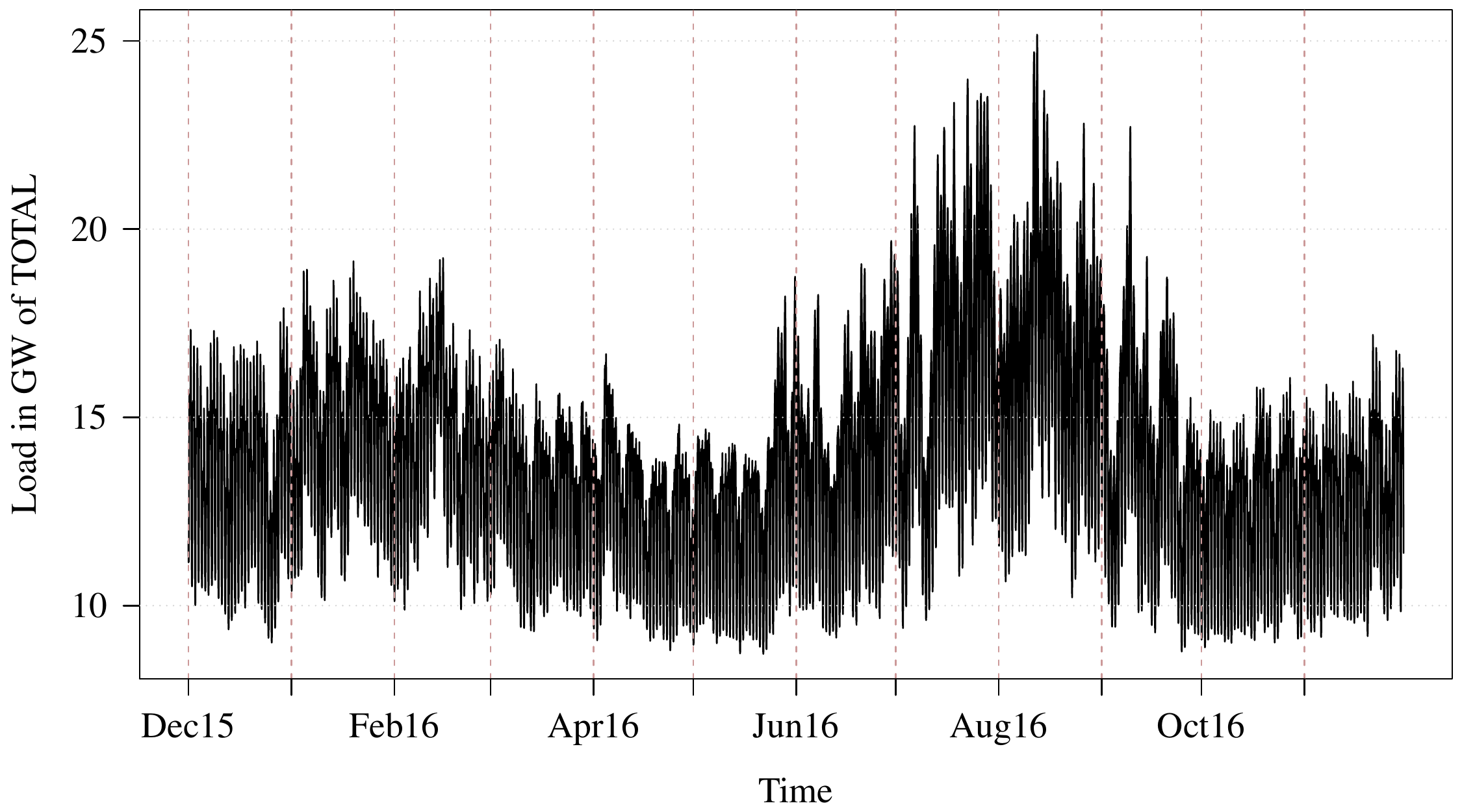} 
\includegraphics[width=.49\textwidth]{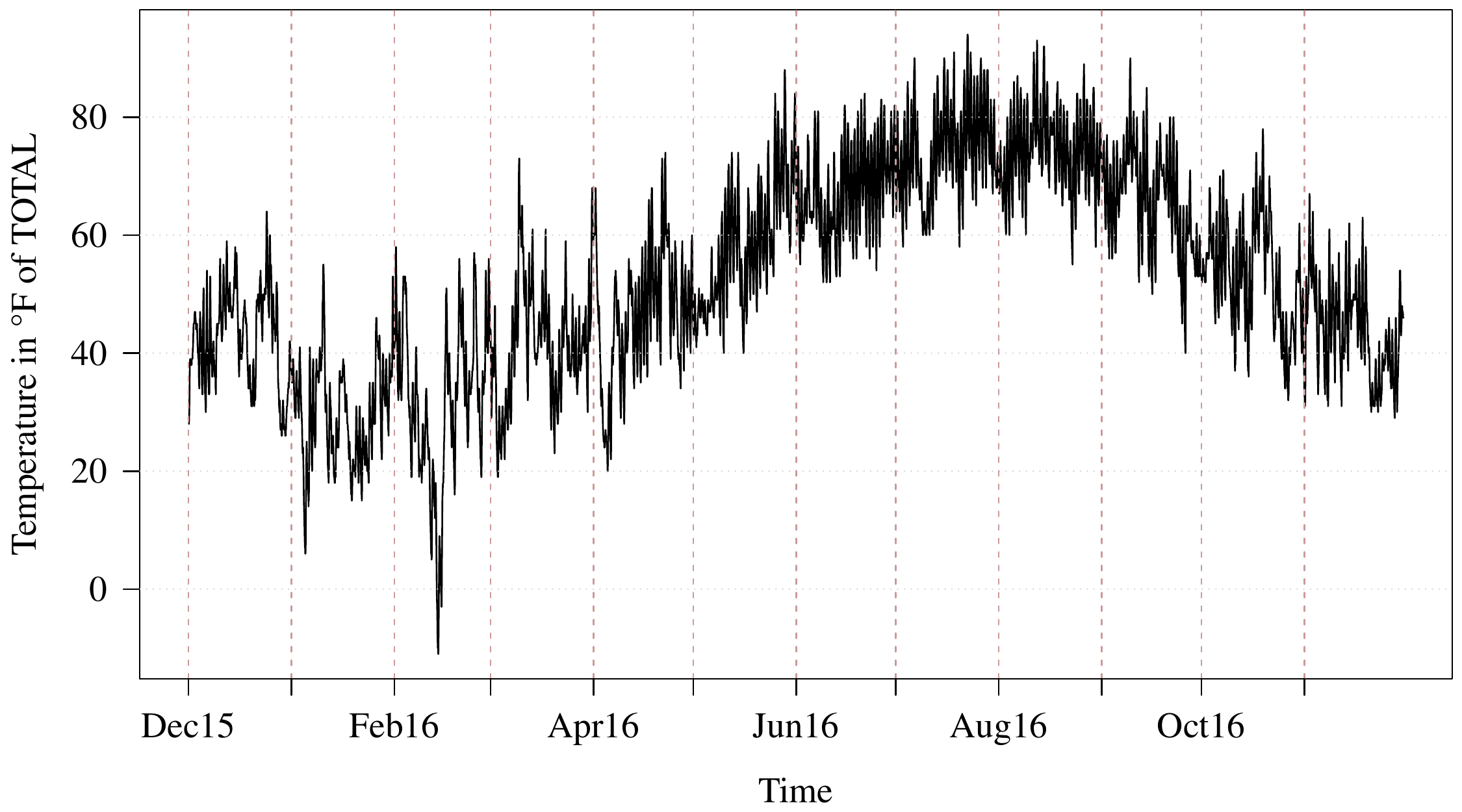} 
\caption{Load (left) and temperature (right) data from Dec 2015 to Nov 2016. This period is the last year of available data for the region TOTAL for the first task.}
\label{fig_data}
\end{figure}

Figure \ref{fig_data} shows the ISO New England zone (TOTAL) load and temperature data for the most recently available year.
The figure depicts clearly a usual annual pattern.
The load level peaks in winter and summer times. The former peak can be explained by higher needs in heating and illumination, the latter by a surge in the demand for air cooling.

We have observed that the data exhibits annual and weekly seasonal patterns. Finally, we want to illustrate that there are so-called interaction effects between both seasonalities.
Essentially, this means that the weekly (including the daily) periodic pattern is changing its shape over the year.
To visualize this effect we transform the load data with respect to annual time periodicity into polar coordinates. Hence, we consider
$$\ell_{\text{x},i,t} = \ell_{i,t} \cos\left( \frac{2\pi t}{24A} \right)\ \ \text{ and } \ \ \ell_{\text{y},i,t} = \ell_{i,t} \sin\left( \frac{2\pi t}{24A} \right) $$
where $24A = 24\times 365.24$ is the (meteorologic) period of a year. 

Figure 4 plots $\ell_{\text{x},i,t}$ and $\ell_{\text{y},i,t}$ and thus visualizes the data in a form analogous to that of a wind rose popular in wind energy forecasting.
The (Euclidean) distance to the center of the figure is the load level of the corresponding load value.
The angle represents the time of the year, the four seasons match the four quadrants. We also highlight 
the hours of the day using different colors to put emphasis on interaction effects.
We observe that the lower load values (those which are closer to the origin) occur during the night hours throughout the year. 
In summer (also late spring and early fall) we sometimes observe very extreme load values, also known as peaks. These peaks can occur at all day hours from around 9am to 9pm. 
In winter (also late fall) we similarly see large load values of around 20GW. However, in contrast to the summer peaks, they always occur during late afternoon and evening hours from 
around 4pm to 9pm. Hence, we can conclude that the structure of the weekly (including the daily) pattern changes over the year. Therefore, we have to take into account 
this effect when developing the forecasting model. 
\begin{figure}[htb!]
\centering
\includegraphics[width=.89\textwidth]{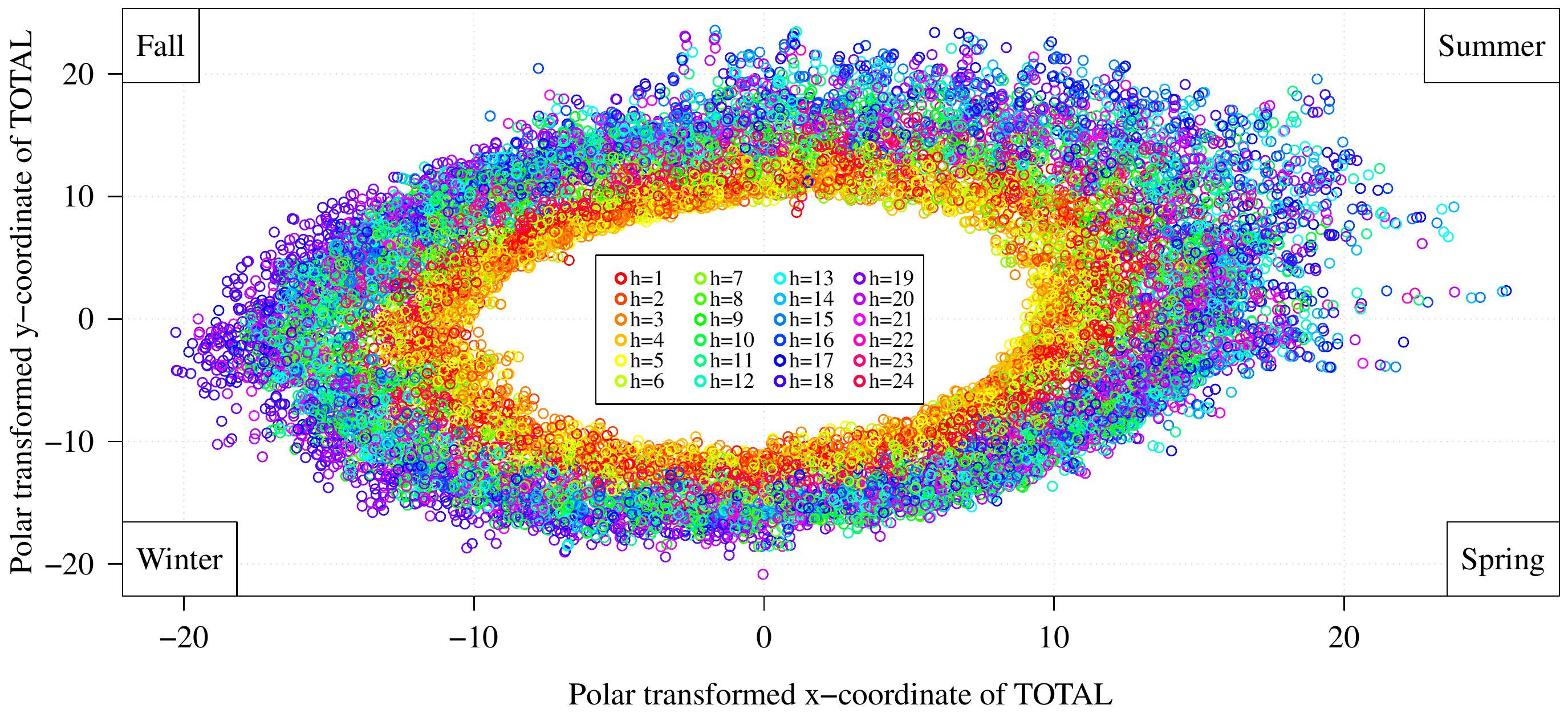} 
\caption{Polar transformed load data $\ell_{\text{x},i,t}$ and $\ell_{\text{y},i,t}$ in GW with respect to annual 
periodicity of TOTAL with hours of the day and seasons marked in different colors}
\label{fig_data_circ}
\end{figure}

Finally, we employ the logarithm of the load data for the main parts of the modeling procedure. Thus, we introduce $L_{i,t} = \log( \ell_{i,t} )$ as log-load where we use MW as original 
unit for $\ell_{i,t}$.

\section{The modeling and forecasting methodology}
 
As mentioned previously, we model each load time series individually, i.e. we ignore the hierarchical structure.
Note that incoherency in the load forecasting regarding the hierarchical structure is not penalized by the evaluation score. More details on coherent probabilistic load forecasts can be found in \cite{taieb2017coherent}. 

Thus, 
for each region $i$, the load $\ell_{i,t}$ (or log-load $L_{i,t}$) is modeled in exactly the same way. Therefore, we drop for convenience the index $i$ in the modeling and forecasting part.

A key element of the considered methodology is the decomposition of $L_t$ into a long-term trend component and a more stable component $Y_t$:
  
  \begin{equation}
 L_t = \text{trend}_t + Y_t  .
  \end{equation}
This transformation is motivated from the stochastic point of view by the fact that $Y_t$ is closer than $L_t$ to a (periodically) stationary process. This simplifies the forthcoming analysis substantially. 
The long term trend component, $\text{trend}_t$, is modeled by a (smoothing) moving average type model. The main term, $Y_t$,
  is described by a simple quantile regression model.
  
  \subsection{The long term trend model}

  The trend component is not just modeled by a simple linear trend $\beta_0 + \beta_1 t$ 
  but by a specific (smoothing) moving average type model.
Therefore, we consider first the linear regression  
 \begin{equation}
  L_t = \bsbeta' \bsX_t + \epsilon_t
\label{eq_trend} 
 \end{equation}
 with $\bsbeta$ as a parameter vector, $\epsilon_t$ as an error term and $\bsX_t$ as a regression matrix containing daily/weekly dummies, annually periodic basis functions 
such as temperature impacts.
Formally, we
have
\begin{equation}
\bsbeta' \bsX_t
= 
\underbrace{ \bsbeta_1 \WW_t 
+ \sum_{j=1}^4 \bsbeta_{1+j} \AA^{\text{FB2}}_{t,j} \WW_t
+ \sum_{j=1}^{11} \bsbeta_{5+j} \AA^{\text{BS12}}_{t,j} \HH_t
}_{\text{periodically deterministic time effects}}
+ 
\underbrace{\bsbeta_{17} \TT_t + \sum_{j=1}^4 \bsbeta_{17+j} \AA^{\text{FB2}}_{t,j} \TT_t}_{\text{temperature effects}} 
\label{eq_trend_X}
\end{equation}
where $\TT_t = (T_t, T_t^2, T_t^3)'$ is a vector with the first three temperature polynomials,
$\HH_t$ is a vector of length $24$ which includes all 24 hour-of-the-day dummies,
$\WW_t$ is a vector of length $24\times 7 = 168$ which contains all 168 hour-of-the-week dummies,
$\AA^{\text{FB2}}_t$ is a Fourier basis with annual periodicity of order 2 and 
$\AA^{\text{BS12}}_t$ is cubic B-spline basis on an equidistant grid with 12 basis functions.
All these components except the temperature part occur in the model description of $Y_t$ too and will be explained there in more detail.
The polynomial temperature structure is assumed in many simple structured models, e.g. in \cite{ziel2016forecasting}. 
We consider a periodic cubic B-spline to benefits from the locality property. However, for $n$
Therefore we consider a mixture of periodic B-splines and Fourier-basis. However, considering only B-splines would have likely given similar results.
 Note that the last (12th) periodic B-spline component  is dropped due to its collinearity with the day of the week dummy component. 
The overall parameter vector is $\bsbeta = (\bsbeta_1,\bsbeta_2,\ldots, \bsbeta_{21})'$ where the vectors $\bsbeta_i$ have fitting
model dimensions.

We estimate the linear regression equation \eqref{eq_trend} by ordinary least squares (OLS) using the given input data.
This yields directly the residuals $\what{\epsilon}_t$ which we use for the estimation of the trend model.

 The model for $\text{trend}_t$ is an annual (smoothing) moving average of $\epsilon_t$ given by 
$$\text{trend}_t = \frac{1}{K}\sum_{k=1}^{K} \epsilon_{t-k} + e_t$$
 with $e_t$ as an error term and a moving average window of about a year, so $K=52\times 7\times24= 8736$. To calculate $K$ we use a multiple of $7\time24$ to avoid potential bias due to the weekly periodicity in $\epsilon_{t}$.
Given the above results, we can easily estimate $\text{trend}_t$ using the plug-in principle: 
$$\what{\text{trend}}_t = \frac{1}{K}\sum_{k=1}^{K} \what{\epsilon}_{t-k}.$$

  \begin{figure}
 
\includegraphics[width=.99\textwidth]{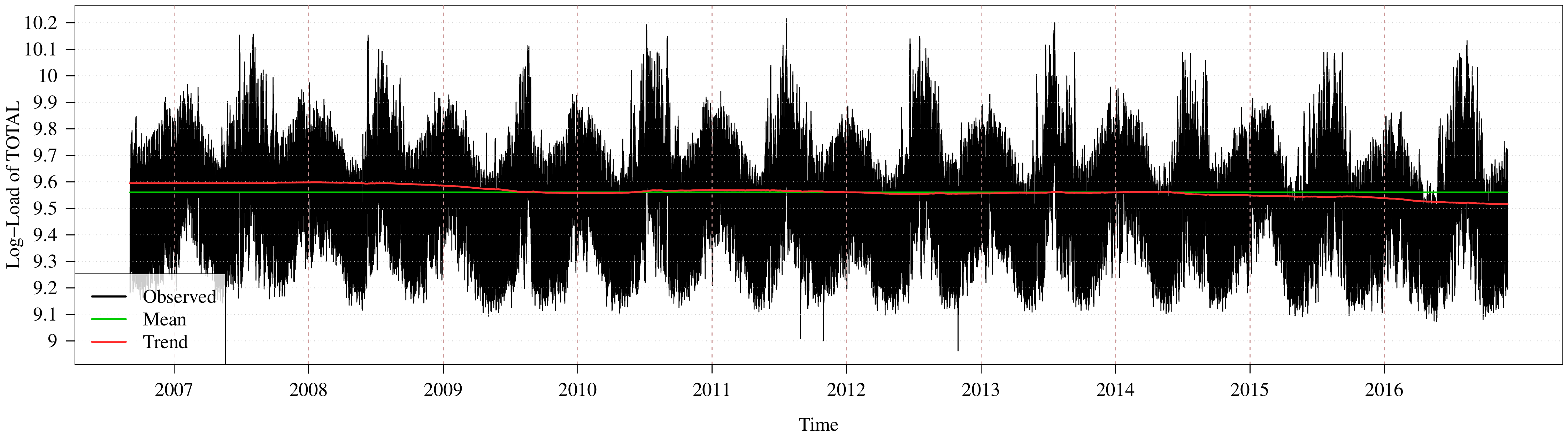} 
\caption{Log-load data $L_t$ (Observed, black) with its sample mean $\what{\mu}$ (Mean, green) and the estimated long-term trend plus sample mean component $\what{\text{trend}}_t + \what{\mu}$ (Trend, red).}
\label{fig_trend}
\end{figure}
\begin{figure} 
\includegraphics[width=.49\textwidth]{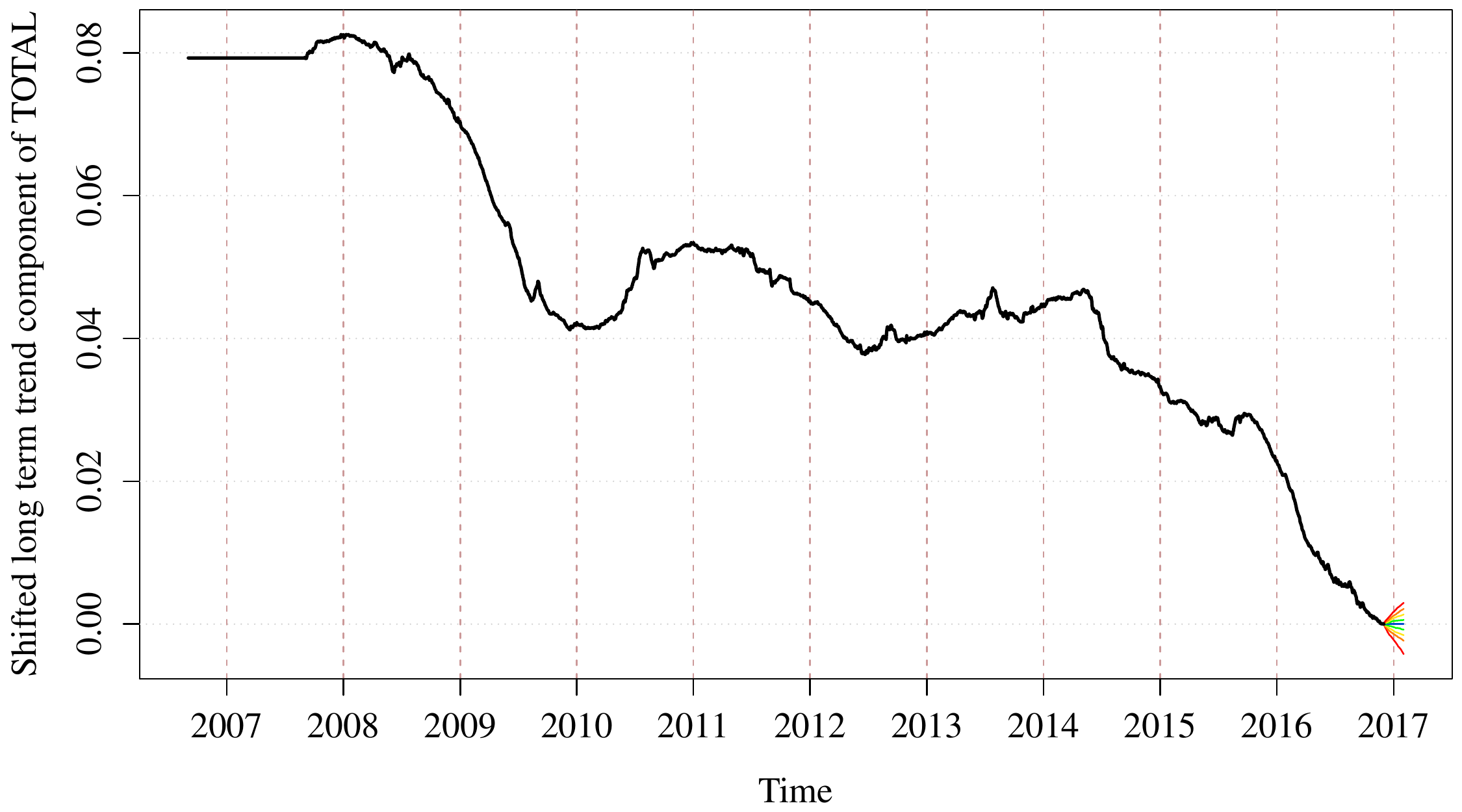} 
\includegraphics[width=.49\textwidth]{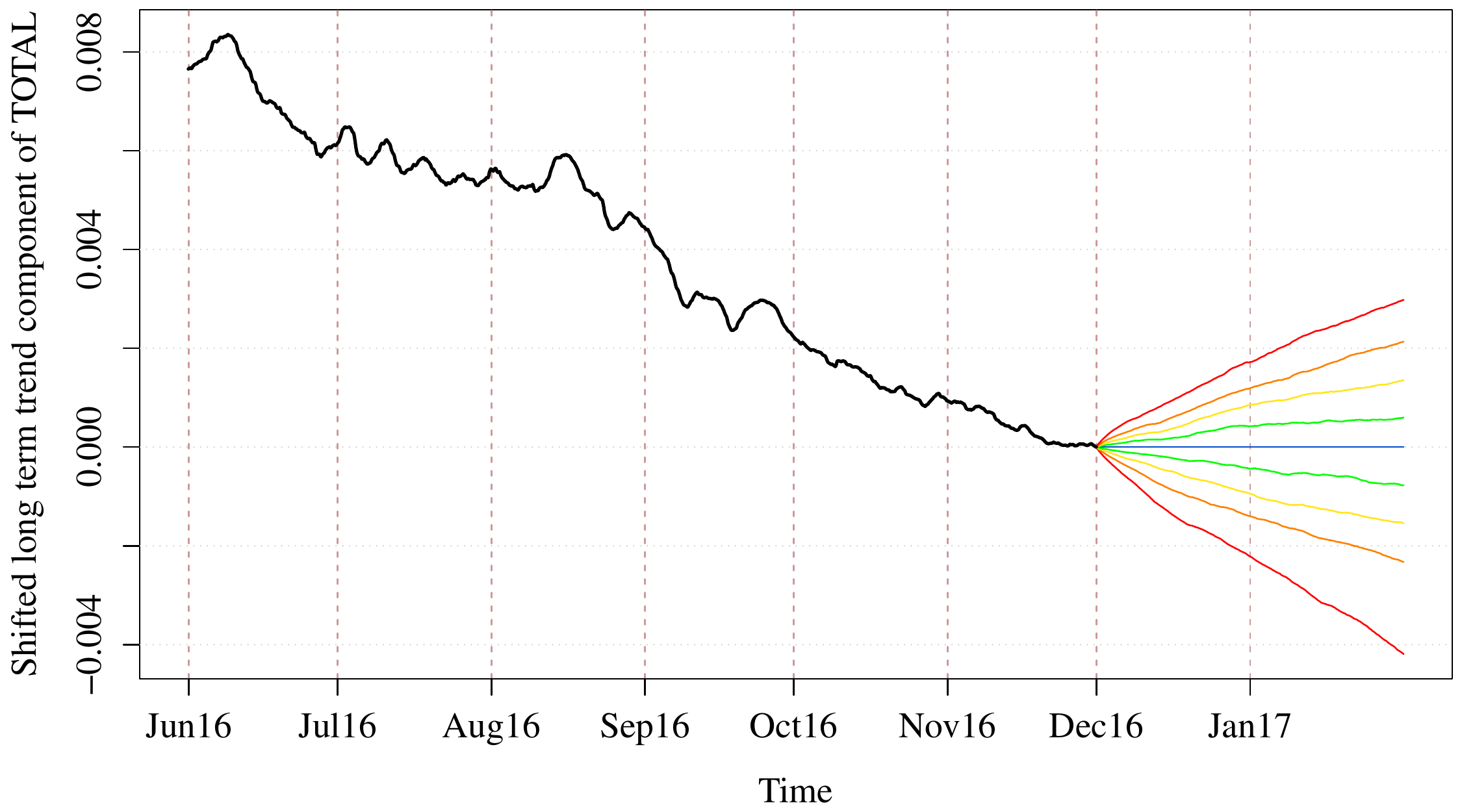} 
\caption{Long-term trend $\what{\text{trend}}_t$ shifted by $\what{\text{trend}}_{t_\text{last}}$ (black) with colored quantile forecasts of all $\tau \in \QQ$ (0.1 and 0.9 red, 0.2 and 0.8 orange, 0.3 and 0.7 yellow, 0.4 and 0.6 green, 
0.5 blue)}
\label{fig_trend_uncert}
\end{figure}

 In Figure \ref{fig_trend} the estimate of the first task is visualized. There we see a downwards trend which is explained by $\what{\text{trend}}_t$.
 Moreover, we can vaguely see that for the first year the trend component is set to be constant (this can be better seen in Figure \ref{fig_trend_uncert}). 
 Nevertheless, this has no impact on the current forecast.

The main purpose of removing the trend component is to create a more stationary type remainder process $Y_t$. If we would do only point forecasting, 
we could stop analyzing $\text{trend}_t$ as its forecast is simply the last known value.
However, here we are interested in probabilistic forecasting. Thus, we have to take into account the uncertainty in the long-term component as well.
To be more precise, we would like to know the quantiles $q_\tau(\text{trend}_{t_{\text{last}}+H   })$ for all $\tau \in \QQ$ at time $t_{\text{last}}+H$ within the out-of-sample range of interest
where $t_{\text{last}}$ is the last in-sample time point and $H$ is the forecasting horizon.
Here, we consider an extremely simple approach to estimate   $q_\tau(\text{trend}_t)$.
We estimate it by taking the sample quantiles for each $\tau \in \QQ$ 
of $\Delta_H \what{\text{trend}}_t = \what{\text{trend}}_t - \what{\text{trend}}_{t-H}$ within the in-sample period but ignore the first $K$ values as they are constant.
Afterwards, we adjust the quantile forecasts by the corresponding median forecasts
and add the last available trend value $\what{\text{trend}}_{t_\text{last}} $, as we assume that the median long-term trend is constant.

The corresponding result of task 1 is given in Figure \ref{fig_trend_uncert} with the full in-sample period and only  the last year of data.
There we see that, compared to the 10 year in-sample period, the long term trend uncertainty is small as the forecasting horizon is relatively short.
We also observe that the median forecast is fixed to be zero.

Note that from the theoretical point of view this approach is sloppy 
as we can not simply add quantile estimates to receive a consistent quantile estimate of $L_t$. 
However, we tested this procedure in out-of-sample test prior to the beginning of the competition and improved the forecasting accuracy in terms of the quantile loss.

\subsection{The quantile regression model} 

For the remainder component $Y_t$ we exploit a simple quantile regression approach.
However, instead of defining a single quantile regression, we consider 24 smaller models for each hour individually.
For a more detailed discussion on this issue see \cite{ziel2016day}. In turn, focusing on 24 models allows us to reduce computational costs and track the hourly specific structures in $Y_t$ more easily.
To simplify notation we introduce the variable $Y_{d,h}$ which corresponds to the $Y_t$ value at hour $h$ and day $d$ (i.e. since $t=24(d-1)+h$). 
 Similarly, this notation is 
 used for all other objects (e.g. $\BB_{d,h}$ for $\BB_{t}$).

Remember that $Y_{d,h}$ carries all non-long term structures in the data, especially the annual, weekly and daily seasonal components.
 Hence, before we turn to the model definition, we denote
 another set of dummies, namely the day-of-the-week dummies. 
Thus, let $\DD_{d}$ be a dummy vector of length $7$ with days of the week as its entries. Note that it 
holds that $\BB_{d,h} = \text{vec}(\HH_{d,h} \otimes \DD_d)$, so the hour-of-the-week dummy basis is the vectorization of the outer product of the hour-of-the-week basis 
and the day-of-the-week basis.

As mentioned before, we consider a quantile regression model. Thus, we define a regression model for the quantile $q_\tau(Y_{d,h})$ for each $\tau\in \QQ$.
So the quantile regression model for $q_\tau(Y_{d,h})$ for each hour $h$ is given by
\begin{equation}
  q_\tau(Y_{d,h}) =
\underbrace{  \bsbeta_{\tau,h,1} \DD_{d} }_{\text{day of the week effects}}
+ 
\underbrace{ \sum_{j=1}^4 \bsbeta_{\tau,h,1+j} \AA^{\text{FB2}}_{d,h,j} \DD_{d} }_{\text{annual effects with day of the week interaction}}
+ 
\underbrace{ \sum_{j=1}^{11} \bsbeta_{\tau,h,5+j} \AA^{\text{BS12}}_{d,h,j} }_{\text{annual effects}}
+ \eps_t
\label{eq_quantreg}
 \end{equation}
 The formula above implies that each quantile regression model has $46$ parameters, as the length of the vector $\bsbeta_{\tau, h} = (\bsbeta_{\tau,h,1}, \ldots , \bsbeta_{\tau,h,16})'$ equals to 46.
 The latter number can be calculated as follows.
 The vector $\bsbeta_{\tau,h,1}$ has the $7$ day-of-the-week dummy elements of $\DD_d$. Then, since each of the vectors
 $\bsbeta_{\tau,h,2}, \ldots, \bsbeta_{\tau,h,5}$ incorporates 7 factors, there are $4\times7=28$ parameters for the annual-weekly interaction effects. Finally, there are
 11 vectors $\bsbeta_{\tau,h,6}, \ldots, \bsbeta_{\tau,h,16}$ of length 1 which account for the annual component. 
 Through the above results it is clear that the number of the model parameters equals to $7+28+11=46$.
 Note that, analogously to the previous case, the last periodic B-spline component  is dropped due to its collinearity with the day of the week dummies.

 However, we observe that only three components in \eqref{eq_quantreg} characterize 
 the full seasonal pattern. The day of the week effects are 
 described by standard day of the week dummies with seven parameters. Thus, for every day of the week we assume a different load level.
 
 The third term in \eqref{eq_quantreg} contains the periodic cubic B-splines with a periodicity of a (meteorologic) year ($A=365.24 \times 24$). 
 This term acts similarly to the monthly dummies but in a smooth manner, i.e. there is no abrupt jump if the month changes.
To be able to achieve a higher performance of the constructed model, it is important that we extend the latter with a periodically smooth basis. 
Naturally, a simple Fourier basis which 
 is based upon sine and cosine functions can do a similar job. Still, the locality property of the B-splines are their advantage.
 Note that the locality property is especially plausible if we would like to consider non-equidistant grids 
 (e.g. to capture more distinct changes within the Christmas and New Years Day periods), but we refrain from doing so in this simple modeling approach.
 For more information on the construction and details of periodic B-splines we recommend \cite{ziel2016lasso} and \cite{ziel2016forecasting}.
 
 Finally, there is the second term in \eqref{eq_quantreg} which models the interaction effect between the day of the week and annual seasonal components.
Hence, this term describes how the weekly pattern is changing 
over the year. To capture this effect we simply multiply the day of the week basis functions $\DD_d$ by
an annual basis functions. Here we choose a Fourier basis of annual periodicity $A$ of order 2 (so it contains 2 sine and cosine waves of period $A$ and $A/2$).
Similarly, to the case before, the periodic B-splines could be applied instead. Still, it is important that the number of parameters (here $4$) which describe the amount of 
annual components in the interaction effect is smaller than or at maximum equals to the plain (non-interactive) annual counterpart.
This condition should be satisfied because the parameter space blows up speedily in response to a hike in the size of the interaction component. 
The correctness of this statement can be illustrated on the example of the current model: more than a  half of all parameters (28 out of 46) come from the 
interaction component. Still, it appears rational 
to have at least four annual basis functions in the interaction component as we observe different effects during all four 
seasons, see Figure \eqref{fig_data_circ}.

  Note that the right hand side of \eqref{eq_quantreg} has the same structure as   \eqref{eq_trend_X} 
  except for the temperature components which are neglected. 
  For many load forecasters it might sound counterintuitive that temperature effects are completely ignored here
  as the influence of temperature on electricity demand 
  is often regarded as very important. However, this is only true for short term forecasting, where we have 
  more accurate weather forecasts available.\footnote{The winners of the GEFCom2014 month ahead probabilistic load forecasting track 
  therefore constructed a separate model approach for the first two days where better temperature information was available, see \cite{goude2014local}.}
 In turn, precision of the forecasts made for a longer horizon is poor.
 Therefore, the temperature carries only a small portion of relevant information for us.
  It is important to remember that all seasonal time-series information can be extracted without using temperature data, for example we can easily infer
  that the electricity demand is higher in winter than in spring just by studying load data (and its derivatives like $Y_{d,h}$).
Because of similar reasons, we do not include any autoregressive components into the model, as we assume that the autoregressive memory is too weak 
on a forecasting horizon of at least a month. This implies that we can not substantially improve the forecasting accuracy by incorporating autoregressive effects.

Given the model specification in \eqref{eq_quantreg}, we can easily estimate the $24\times 9 \times 10$ (hours$\times$quantile levels$\times$zones) quantile regressions.
Here the parameters $\bsbeta_{\tau, h}$ are optimized such that that the pinball score $S_\tau$ in \eqref{eq_pb} (or quantile loss) 
of $\bsbeta_{\tau, h}'\bsX_{d,h}$ becomes minimal with respect to $Y_{d,h}$ over the full in-sample period. Formally, we get 
the quantile estimator by minimizing the following expression
\begin{equation}
 \what{\bsbeta}_{\tau,h} = \argmin_{\bsbeta} \sum_{d\in \DD_{\text{InSample}}} S_\tau( Y_{d,h} , \bsbeta' \bsX_{d,h})
\label{eq_quantreg_argmin}
 \end{equation}
where $\DD_{\text{InSample}}$ is the set of all in-sample days.
We estimate the coefficients in \eqref{eq_quantreg_argmin} by noting that the minimization problem in \eqref{eq_quantreg_argmin} is equivalent to maximum likelihood estimation of the linear regression problem 
$Y_{d,h} = \bsbeta_h' \bsX_{d,h} + \eps_{d,h}$ 
where the error term $\eps_{d,h}$ is assumed to follow an asymmetric Laplace distribution with skewness parameter $\tau$.

\begin{figure} 
\includegraphics[width=.99\textwidth]{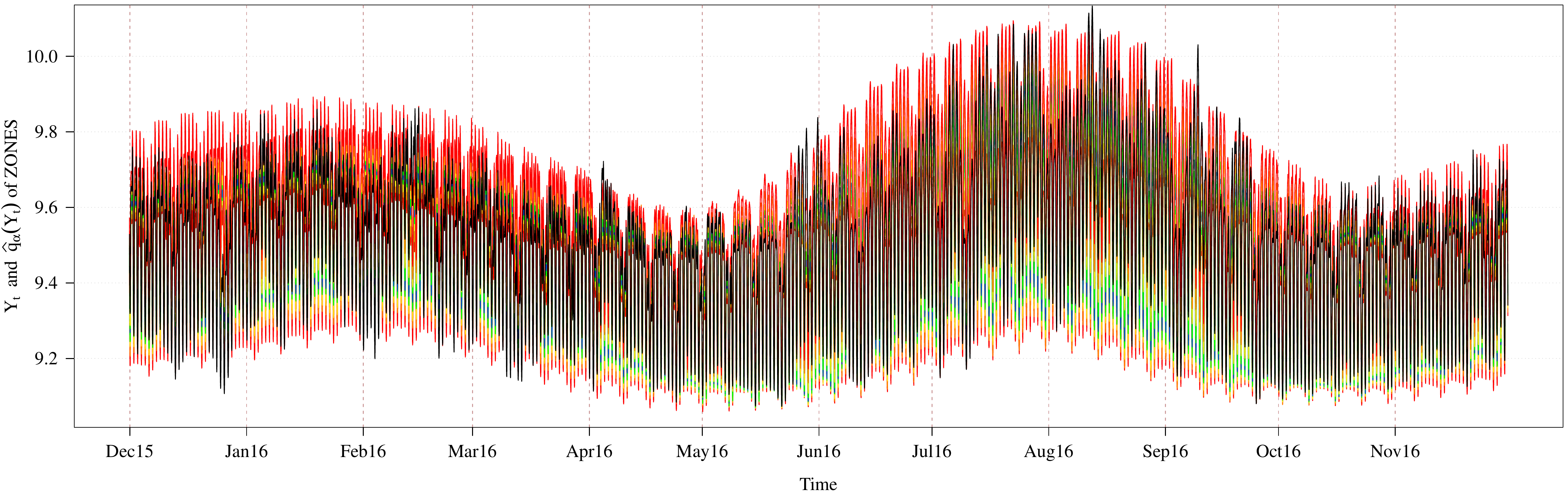} 
\includegraphics[width=.99\textwidth]{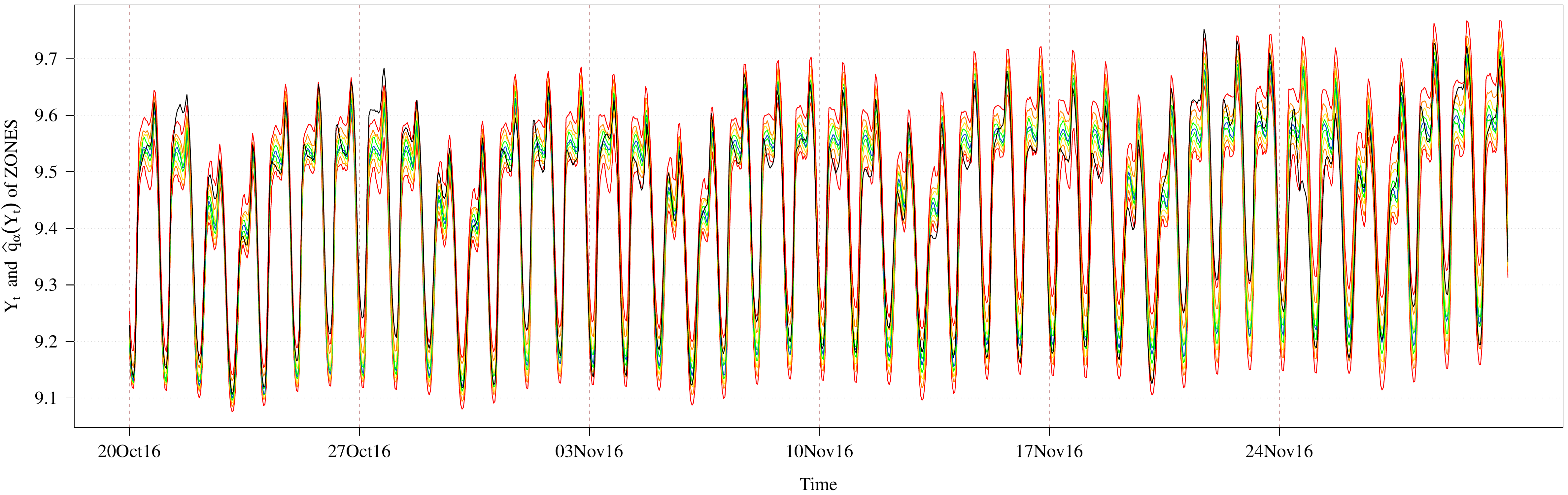} 
\caption{Observed log-load (black) with colored quantile fits of all quantiles $\tau \in \QQ$ (0.1 and 0.9 red, 0.2 and 0.8 orange, 0.3 and 0.7 yellow, 0.4 and 0.6 green, 
0.5 blue) for the most recent year and the most recent six weeks of ZONES.}
\label{fig_quantile_insample}
\end{figure}

For the implementation the outlined model above we consider the \texttt{R}-package \texttt{quantreg} (see \cite{quantregRpackage}) with default inputs.  
Figure \ref{fig_quantile_insample} shows the in-sample fit of the most recent year and the most recent six weeks of the assumed quantile model (referred to as ZONES).
We see that, in general, the fit seems to capture not only the annual but also the weekly seasonal effects well. Also, the seasonal interaction effect is visible: 
In October the forecast of the magnitude of the evening peak is much less distinct than at  the end of November, so the weekly pattern changes smoothly and appreciably over the six weeks.
The figure also demonstrates that the range of the 10\% quantile to 90\% quantile is roughly 0.1. 
In contrast, the corresponding 10\%-90\% quantile range for the 
two month ahead forecast of $\text{trend}_t$ is less than 0.008, see Figure \ref{fig_trend_uncert}. Thus, most of the uncertainty in the forecast will be driven by the quantile regression component $Y_{d,h}$.

Using the estimated parameter vector $\what{\bsbeta}_{\tau,h}$ we can compute the quantile forecasts of $Y_{d,h}$ for each model by 
evaluating $\what{\bsbeta}_{\tau,h}' \bsX_t$ for time point $t$.
As all regressors in $\bsX_t$ are deterministic we can directly compute the forecasts with a forecasting horizon of $D$ days. These forecasts are denoted by
$\what{q}_{\tau}(Y_{d_{\text{last}}+D, h})$. In turn, the latter expression can be index converted to $\what{q}_{\tau}(Y_{t_{\text{last}}+H})$\footnote{The forecasting horizon $H$ can be written as $H=24(D-1)+h$.}.
Afterwards, we estimate the final quantile log-load forecast by 
adding the quantile forecasts of $Y_t$ and $\text{trend}_t$ along the formula
$\what{q}_{\tau}( L_{t_{\text{last}} + H} ) = \what{q}_{\tau}(Y_{t_{\text{last}}+H}) + \what{q}_{\tau}(\text{trend}_{t_{\text{last}}+H})  $. 
As mentioned before, this summation of quantiles  can be ambiguous (it is only correct if $\text{trend}_t$ and $Y_t$ are comonotone), but the method seems to work sufficiently well in application.
This was also tested in preliminary back-testing studies prior to the start of the competition.
Finally, quantiles are preserved by strictly monotonic transforms and the 
exponential function $\exp$ as inverse of the logarithm $\log$ is strictly increasing. Thus, we obtain the quantile forecast of the load $\ell_t$ given by
$\what{q}_{\tau}( \ell_{t_{\text{last}} + H} ) = \exp( \what{q}_{\tau}( L_{t_{\text{last}} + H} )  )$.

\begin{figure} 
\includegraphics[width=.99\textwidth]{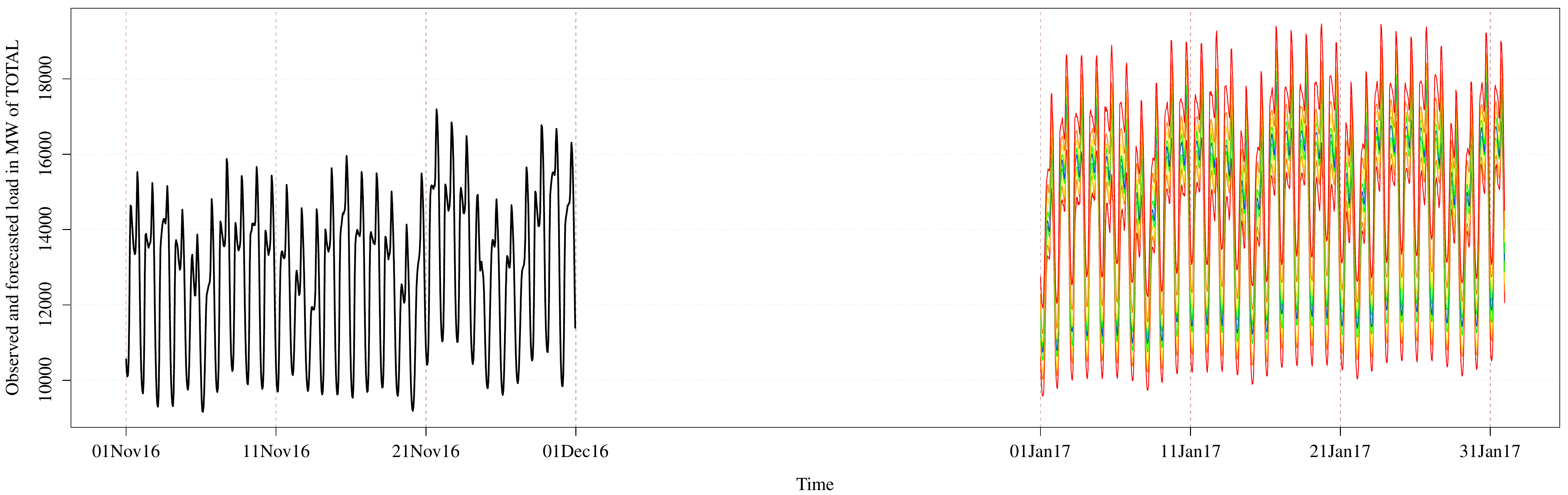} 
\caption{Observed load (black) and forecasted load (colored) for all quantiles $\tau \in \QQ$ (0.1 and 0.9 red, 0.2 and 0.8 orange, 0.3 and 0.7 yellow, 0.4 and 0.6 green, 
0.5 blue) for the first task.}
\label{fig_quantile_forecast}
\end{figure}

Figure \ref{fig_quantile_forecast} shows the forecast of the ISO New England zone for the first task. 
As the most recently available data was that at the end of November this is a two month ahead forecast. 
However, only the forecast for the latter month is submitted and evaluated.
To emphasize this and visualize the the obtained results we removed the forecast values for January.
We observe that the overall load level increased from November to January as we would expect it due to  a respective drop in average daily temperature. 
However, the forecasted weekly pattern seem to be quite stable over January. Nevertheless, in contrast to Figure \ref{fig_quantile_insample} we observe a small tendency
that the evening peak decreases slightly at the end of January. This can be possibly explained by the fact that temperatures are starting to become warmer and the duration of the daylight increases. 
As a consequence, less electricity is being demanded.

\section{Results and Conclusion}

The proposed methodology allowed us to take 2nd place in the open data track and 4th place in the definite data track. 
In both tracks, only five teams succeed to achieve this excellent level of performance. 

In Table \ref{tab_results} we show the overall results in terms of the pinball score for each task and each zone for the proposed model.
There we also give the results of the Vanilla Benchmark (fore more details on the Vanilla Benchmark see \cite{hong2016probabilistic}) which is used in the final evaluation.
Next to the pinball scores we highlight in Table \ref{tab_results} the improvement in the pinball score with respect to the benchmark.
We see that in 90\% of all cases the pinball score is smaller than the Vanilla benchmark.
Given the simplicity of the model we developed, its forecasting accuracy appears
remarkable and shows a high potential of the quantile regression methodology.


\begin{table}[ht]
\centering
\tabcolsep=0.12cm
\begin{tabular}{|r|r|rrrrrrrrrr|}
  \hline
 \#& \small PB Score& \small TOTAL & \small ME & \small NH & \small VT & \small CT & \small RI & \small MASS & \small MA.SE & \small MA.WC & \small MA.NE \\ 
  \hline
 &Model (MW) & 358.44 & 23.33 & 35.97 & 20.69 & 106.82 & 22.38 & 160.72 & 43.24 & 49.56 & 71.01 \\ 
  \textbf{1}& Benchmark (MW) & 402.68 & 36.95 & 41.91 & 22.44 & 114.88 & 23.32 & 170.20 & 44.11 & 50.58 & 77.85 \\ 
  & Improvement (\%)& $ \cellcolor[rgb]{0.482,0.891,0.391} 10.99$ & $ \cellcolor[rgb]{0.361,0.831,0.331} 36.87$ & $ \cellcolor[rgb]{0.467,0.884,0.384} 14.16$ & $ \cellcolor[rgb]{0.497,0.898,0.398} 7.79$ & $ \cellcolor[rgb]{0.509,0.902,0.402} 7.02$ & $ \cellcolor[rgb]{0.719,0.944,0.444} 4.01$ & $ \cellcolor[rgb]{0.61,0.922,0.422} 5.57$ & $ \cellcolor[rgb]{0.861,0.972,0.472} 1.98$ & $ \cellcolor[rgb]{0.858,0.972,0.472} 2.02$ & $ \cellcolor[rgb]{0.492,0.896,0.396} 8.78$ \\ \hline
  & Model (MW) & 369.05 & 22.72 & 31.65 & 14.91 & 111.28 & 24.56 & 179.17 & 48.75 & 58.29 & 75.29 \\ 
  \textbf{2}& Benchmark (MW) & 401.51 & 29.11 & 35.34 & 15.49 & 115.72 & 24.18 & 190.36 & 50.69 & 60.32 & 81.02 \\ 
  & Improvement (\%) & $ \cellcolor[rgb]{0.496,0.898,0.398} 8.08$ & $ \cellcolor[rgb]{0.431,0.865,0.365} 21.95$ & $ \cellcolor[rgb]{0.485,0.892,0.392} 10.44$ & $ \cellcolor[rgb]{0.736,0.947,0.447} 3.77$ & $ \cellcolor[rgb]{0.732,0.946,0.446} 3.83$ & $ \cellcolor[rgb]{1,0.922,0.5} -1.55$ & $ \cellcolor[rgb]{0.589,0.918,0.418} 5.88$ & $ \cellcolor[rgb]{0.732,0.946,0.446} 3.83$ & $ \cellcolor[rgb]{0.764,0.953,0.453} 3.37$ & $ \cellcolor[rgb]{0.505,0.901,0.401} 7.07$ \\ \hline
  & Model (MW) & 369.05 & 22.72 & 31.65 & 14.91 & 111.28 & 24.56 & 179.17 & 48.75 & 58.29 & 75.29 \\ 
  \textbf{3}& Benchmark (MW) & 401.51 & 29.11 & 35.34 & 15.49 & 115.72 & 24.18 & 190.36 & 50.69 & 60.32 & 81.02 \\ 
  & Improvement (\%) & $ \cellcolor[rgb]{0.496,0.898,0.398} 8.08$ & $ \cellcolor[rgb]{0.431,0.865,0.365} 21.95$ & $ \cellcolor[rgb]{0.485,0.892,0.392} 10.44$ & $ \cellcolor[rgb]{0.736,0.947,0.447} 3.77$ & $ \cellcolor[rgb]{0.732,0.946,0.446} 3.83$ & $ \cellcolor[rgb]{1,0.922,0.5} -1.55$ & $ \cellcolor[rgb]{0.589,0.918,0.418} 5.88$ & $ \cellcolor[rgb]{0.732,0.946,0.446} 3.83$ & $ \cellcolor[rgb]{0.764,0.953,0.453} 3.37$ & $ \cellcolor[rgb]{0.505,0.901,0.401} 7.07$ \\ \hline
  & Model (MW) & 346.68 & 23.52 & 30.37 & 20.33 & 95.81 & 20.34 & 173.63 & 47.63 & 53.84 & 73.48 \\ 
  \textbf{4}& Benchmark (MW) & 351.89 & 23.96 & 29.43 & 21.07 & 98.91 & 21.54 & 175.86 & 49.62 & 55.43 & 73.32 \\ 
  & Improvement (\%) & $ \cellcolor[rgb]{0.896,0.979,0.479} 1.48$ & $ \cellcolor[rgb]{0.872,0.974,0.474} 1.83$ & $ \cellcolor[rgb]{1,0.841,0.5} -3.18$ & $ \cellcolor[rgb]{0.755,0.951,0.451} 3.50$ & $ \cellcolor[rgb]{0.78,0.956,0.456} 3.14$ & $ \cellcolor[rgb]{0.609,0.922,0.422} 5.59$ & $ \cellcolor[rgb]{0.911,0.982,0.482} 1.27$ & $ \cellcolor[rgb]{0.719,0.944,0.444} 4.01$ & $ \cellcolor[rgb]{0.8,0.96,0.46} 2.86$ & $ \cellcolor[rgb]{1,0.989,0.5} -0.22$ \\ \hline
  & Model (MW) & 346.68 & 23.52 & 30.37 & 20.33 & 95.81 & 20.34 & 173.63 & 47.63 & 53.84 & 73.48 \\ 
  \textbf{5}& Benchmark (MW) & 351.70 & 23.88 & 29.64 & 20.92 & 98.80 & 21.53 & 175.86 & 49.51 & 55.25 & 73.16 \\ 
  & Improvement (\%) & $ \cellcolor[rgb]{0.9,0.98,0.48} 1.43$ & $ \cellcolor[rgb]{0.895,0.979,0.479} 1.50$ & $ \cellcolor[rgb]{1,0.878,0.5} -2.45$ & $ \cellcolor[rgb]{0.804,0.961,0.461} 2.81$ & $ \cellcolor[rgb]{0.788,0.958,0.458} 3.03$ & $ \cellcolor[rgb]{0.612,0.922,0.422} 5.54$ & $ \cellcolor[rgb]{0.911,0.982,0.482} 1.27$ & $ \cellcolor[rgb]{0.734,0.947,0.447} 3.80$ & $ \cellcolor[rgb]{0.822,0.964,0.464} 2.55$ & $ \cellcolor[rgb]{1,0.978,0.5} -0.44$ \\ \hline
  & Model (MW) & 180.12 & 15.78 & 15.67 & 14.45 & 49.05 & 10.34 & 93.20 & 30.01 & 31.35 & 37.07 \\ 
  \textbf{6}& Benchmark (MW) & 202.83 & 29.71 & 16.74 & 17.23 & 55.11 & 11.19 & 106.5 & 34.19 & 34.91 & 44.41 \\ 
  & Improvement (\%) & $ \cellcolor[rgb]{0.481,0.891,0.391} 11.19$ & $ \cellcolor[rgb]{0.315,0.807,0.307} 46.89$ & $ \cellcolor[rgb]{0.551,0.91,0.41} 6.42$ & $ \cellcolor[rgb]{0.458,0.879,0.379} 16.12$ & $ \cellcolor[rgb]{0.482,0.891,0.391} 11.00$ & $ \cellcolor[rgb]{0.498,0.899,0.399} 7.60$ & $ \cellcolor[rgb]{0.475,0.888,0.388} 12.49$ & $ \cellcolor[rgb]{0.476,0.888,0.388} 12.24$ & $ \cellcolor[rgb]{0.486,0.893,0.393} 10.21$ & $ \cellcolor[rgb]{0.456,0.878,0.378} 16.53$ \\ 
   \hline
\end{tabular}
\caption{Pinball (PB) scores of the considered model and the benchmark with corresponding improvement for all 6 tasks for all 10 zones.}
\label{tab_results}
\end{table}

More advanced effects, e.g. non-linear effects, public holidays effects or structural changes, were ignored in the present model. 
Incorporating structural changes in the long term trend appears to be the most prospective component which would likely boost the accuracy of a model. 
To show the correctness of the previous statement explicitly, consider, for example, that 
Massachusetts had a strong increase in solar power production in recent years, however, not as strong as the one recorded in Connecticut.
This implies that not only the long term trend effect  but also other model components vary over time. This can be seen especially clearly in summer periods, when the load volatility 
tends to be higher during the day hours due to stronger and less predictable 
meteorological impacts. Hence, including these effects would very likely improve the results substantially. 
Furthermore, we could also take into account the load time-series of other zones than the one we are actually forecasting, especially using the neighboring and hierarchical structure information 
might improve the forecasting performance. 

Finally, there is always the \emph{curse of dimensionality} in empirical data analytics to be acknowledged. Even though the considered models are still relatively low-dimensional, shrinkage 
algorithms are to be used to avoid the danger of overfitting. For quantile regression the quantile lasso is a plausible option to be applied in future. 
The \texttt{R}-package \texttt{quantreg} provides such an implementation. 

\section*{References}

\bibliographystyle{apalike}
\bibliography{biblio}

\begin{thebibliography}{}

\bibitem[Gaillard et~al., 2016]{gaillard2016additive}
Gaillard, P., Goude, Y., and Nedellec, R. (2016).
\newblock Additive models and robust aggregation for gefcom2014 probabilistic
  electric load and electricity price forecasting.
\newblock {\em International Journal of Forecasting}, 32(3):1038--1050.

\bibitem[Goude et~al., 2014]{goude2014local}
Goude, Y., Nedellec, R., and Kong, N. (2014).
\newblock Local short and middle term electricity load forecasting with
  semi-parametric additive models.
\newblock {\em IEEE transactions on smart grid}, 5(1):440--446.

\bibitem[Haben and Giasemidis, 2016]{Haben20161017}
Haben, S. and Giasemidis, G. (2016).
\newblock {A hybrid model of kernel density estimation and quantile regression
  for GEFCom2014 probabilistic load forecasting}.
\newblock {\em International Journal of Forecasting}, 32(3):1017--1022.

\bibitem[Hong et~al., 2016]{hong2016probabilistic}
Hong, T., Pinson, P., Fan, S., Zareipour, H., Troccoli, A., and Hyndman, R.~J.
  (2016).
\newblock Probabilistic energy forecasting: Global energy forecasting
  competition 2014 and beyond.
\newblock {\em International Journal of Forecasting}, 32(2):896--913.

\bibitem[Koenker, 2018]{quantregRpackage}
Koenker, R. (2018).
\newblock {\em quantreg: Quantile Regression}.
\newblock R package version 5.35.

\bibitem[Liu et~al., 2017]{liu2017probabilistic}
Liu, B., Nowotarski, J., Hong, T., and Weron, R. (2017).
\newblock Probabilistic load forecasting via quantile regression averaging on
  sister forecasts.
\newblock {\em IEEE Transactions on Smart Grid}, 8(2):730--737.

\bibitem[Maciejowska and Nowotarski, 2016]{maciejowska2016hybrid}
Maciejowska, K. and Nowotarski, J. (2016).
\newblock A hybrid model for gefcom2014 probabilistic electricity price
  forecasting.
\newblock {\em International Journal of Forecasting}, 32(3):1051--1056.

\bibitem[Taieb et~al., 2016]{taieb2016forecasting}
Taieb, S.~B., Huser, R., Hyndman, R.~J., and Genton, M.~G. (2016).
\newblock Forecasting uncertainty in electricity smart meter data by boosting
  additive quantile regression.
\newblock {\em IEEE Transactions on Smart Grid}, 7(5):2448--2455.

\bibitem[Taieb et~al., 2017]{taieb2017coherent}
Taieb, S.~B., Taylor, J.~W., and Hyndman, R.~J. (2017).
\newblock Coherent probabilistic forecasts for hierarchical time series.
\newblock In {\em International Conference on Machine Learning}, pages
  3348--3357.

\bibitem[Ziel et~al., 2016]{ziel2016forecasting}
Ziel, F., Croonenbroeck, C., and Ambach, D. (2016).
\newblock Forecasting wind power--modeling periodic and non-linear effects
  under conditional heteroscedasticity.
\newblock {\em Applied Energy}, 177:285--297.

\bibitem[Ziel and Liu, 2016]{ziel2016lasso}
Ziel, F. and Liu, B. (2016).
\newblock Lasso estimation for gefcom2014 probabilistic electric load
  forecasting.
\newblock {\em International Journal of Forecasting}, 32(3):1029--1037.

\bibitem[Ziel and Weron, 2018]{ziel2016day}
Ziel, F. and Weron, R. (2018).
\newblock Day-ahead electricity price forecasting with high-dimensional
  structures: Univariate vs. multivariate modeling frameworks.
\newblock {\em Energy Economics}, 70:396--420.

\end{thebibliography}

\end{document}